\documentclass{aa}  
\usepackage{graphicx}
\usepackage[varg]{txfonts}
\usepackage{lscape}
\begin{document}

\title{The frequency of planetary debris \\ around young white dwarfs}

   \author{D. Koester \inst{1} \and B.T. G\"ansicke\inst{2} \and
     J. Farihi \inst{3} } 
     \institute{Institut f\"ur Theoretische
     Physik und Astrophysik, Universit\"at Kiel, 24098 Kiel,
     Germany\\ 
     \email{koester@astrophysik.uni-kiel.de} 
     \and Department
     of Physics, University of Warwick, Coventry CV4 7AL, UK 
     \and
     University College London, Department of Physics \& Astronomy,
     Gower Street, London WC1E 6BT, UK }

   \date{submitted 21 Feb 2014}

\abstract {Heavy metals in the atmospheres of white dwarfs are thought in many
  cases to be accreted from a circumstellar debris disc, which was formed by
  the tidal disruption of a rocky planetary body within the Roche radius of
  the star. The abundance analysis of photospheric elements and conclusions
  about the chemical composition of the accreted matter are a new and
  promising method of studying the composition of extrasolar planetary
  systems. However, ground-based searches for metal-polluted white dwarfs that
  rely primarily on the detection of the \ion{Ca}{ii}~K line become
  insensitive at $T_\mathrm{ eff}>15\,000$~K because this ionization state
  depopulates.}  
{ We present the results of the first unbiased survey for metal pollution
  among hydrogen-atmosphere (DA type) white dwarfs with cooling ages in the
  range 20--200~Myr and $17\,000$~K $ < T_\mathrm{eff} < 27\,000$~K.}  
{ The sample was observed with the Cosmic Origins Spectrograph on board the
  Hubble Space Telescope in the far ultraviolet range between 1130 and
  1435\,\AA. The atmospheric parameters were obtained using these spectra and
  optical observations from the literature. Element abundances were determined
  using theoretical models, which include the effects of element
  stratification due to gravitational settling and radiative levitation.}  
{ We find 48 of the 85 DA white dwarfs studied, or 56\% show traces of heavy
  elements. In 25 stars (showing only Si and occasionally C), the elements can
  be explained by radiative levitation alone, although we argue that accretion
  has very likely occurred recently.  The remaining 23 white dwarfs (27\%),
  however, must be currently accreting. Together with previous studies from
  the ground and adopting bulk Earth abundances for the debris, accretion
  rates range from a few $10^5$~g\,s$^{-1}$ to a few $10^8$~g\,s$^{-1}$, with
  no evident trend in cooling age from $\approx$40~Myr to $\approx$2~Gyr. Only
  a single, modest case of metal pollution ($\dot{M} < 10^6$~g\,s$^{-1}$) is
  found among ten white dwarfs with $T_\mathrm{eff} > 23\,000$~K, in excellent
  agreement with the absence of infrared excess from dust around these warmer
  stars.  The median, main sequence progenitor of our sample corresponds to an
  A-type star of $\approx$2~$M_\sun$, and we find 13 of 23 white dwarfs
  descending from main sequence 2--3~$M_\sun$, late B- and A-type stars to be
  currently accreting.  Only one of 14 targets with $M_\mathrm{wd}>0.8~M_\sun$
  is found to be currently accreting, which suggests a large fraction of these
  stars result from double-degenerate mergers, and the merger discs do not
  commonly reform large planetesimals or otherwise pollute the remnant. We
  reconfirm our previous finding that two 625~Myr Hyades white dwarfs are
  currently accreting rocky planetary debris.}  
{At least 27\% of all white dwarfs with cooling ages 20--200~Myr are accreting
  planetary debris, but that fraction could be as high as $\approx$50\%. At
  $T_\mathrm{ eff}>23\,000$~K, the luminosity of white dwarfs is probably
  sufficient to vaporize circumstellar dust grains, so no stars with strong
  metal-pollution are found. Planetesimal disruption events should occur in
  this cooling age and temperature range as well, and they are likely to
  result in short phases of high mass-transfer rates. It appears that the
  formation of rocky planetary material is common around 2--3~$M_\sun$ late B-
  and A-type stars.}


   \maketitle
%

\section{Introduction}
It is becoming increasingly clear that the presence of planets is
more the norm than the exception. Analysis of transiting planet
candidates from \textit{Kepler} puts the frequency of earth-sized
planets with orbital periods $<85$~days at $16.5\pm3.6\%$
\citep{fressin13-1}, whereas microlensing, which is sensitive to
planets on longer period orbits, suggests that the fraction of stars
with cool Neptunes or super Earths is $\approx$50-60\%
\citep{cassanetal12-1}.

A fundamental question that has, until recently, received surprisingly
little attention is the ultimate fate of planetary
  systems once their host stars evolve off the main sequence?.
Initially focusing on the future of the solar system
\citep{Sackmann.Boothroyd.ea93, duncan+lissauer98-1}, a number of
theoretical studies have shown that a fraction of planets can survive
the red-giant stage of their host stars \citep{villaver+livio07-1,
  villaver+livo09-1, nordhausetal10-1, mustill+villaver12-1}.  The
ensuing long-term orbital evolution is complex and may lead to planet
ejections or collisions \citep{Debes.Sigurdsson02, verasetal11-1,
  voyatzisetal13-1}. Smaller bodies are likely to be scattered, from
locations comparable to the solar system's main asteroid belt or
Kuiper belt \citep{bonsoretal11-1, Debes.Walsh.ea12}, which will lead
to their tidal disruption if their trajectory takes them within the
Roche radius of the white dwarf, $\approx$1~$R_\sun$, \citep{davidsson99-1}.

Accretion of debris from the disruption of planetary bodies is now the
canonical explanation \citep{Graham.Matthews.ea90, Jura03} for the
presence of dusty and gaseous discs discovered around 30 white
dwarfs \citep[e.g.,][]{Zuckerman.Becklin87, Becklin.Farihi.ea05,
  Gansicke.Marsh.ea06, Farihi.Zuckerman.ea08, Dufour.Kilic.ea10,
  kilicetal12-1} and the presence of photospheric metals in a large
number of white dwarfs.

The strong surface gravity of white dwarfs leads to gravitational settling:
all heavy elements sink out of the atmosphere with only the lightest one
floating on top \citep{Schatzman47}. This explains the composition of the vast
majority of objects with almost pure hydrogen or helium surfaces. Metals are
expected at the hottest temperatures because radiative levitation
\citep{Michaud.Martel.ea79, Vennes.Pelletier.ea88, Chayer.Fontaine.ea95,
  Chayer.Vennes.ea95}, or at the cool end of the cooling sequence through
convective mixing with deeper layers \citep{Koester.Weidemann.ea82,
  Fontaine.Villeneuve.ea84, Pelletier.Fontaine.ea86}. Nevertheless, small
traces of heavy metals are also found in the intermediate temperature range
\citep[see, e.g.,][for a DA sample]{Koester.Wilken06}, where they can only be
supplied by an external source through accretion. Historically, accretion from
interstellar matter was considered as the most likely explanation
\citep{Fontaine.Michaud79, Vauclair.Vauclair.ea79, Alcock.Illarionov80,
  Dupuis.Fontaine.ea92}. However, this scenario had several problems
\citep[e.g.][]{Aannestad.Kenyon.ea93, Wolff.Koester.ea02,
  Friedrich.Jordan.ea04, Farihi.Barstow.ea10}, and it is now clear that
accretion of planetary debris is the most likely explanation for the majority
of, if not all, metal-polluted white dwarfs.

The direct detection of planetary material in the photosphere of a
white dwarf offers the opportunity to study the chemical composition
of exoplanetary systems with a scope and accuracy that will not be
reached by other methods in the foreseeable future
\citep{Zuckerman.Koester.ea07}. Because high resolution and high
signal-to-noise spectra are needed, detailed studies have so far been
carried out only for a handful of objects
\citep[e.g.][]{Klein.Jura.ea10, Klein.Jura.ea11, Vennes.Kawka.ea11,
  Zuckerman.Koester.ea11, Dufour.Kilic.ea12, Gansicke.Koester.ea12,
  juraetal12-1, Xu.Jura.ea13}, with the common conclusion that the
parent bodies of the accreted debris are rocky. 

The stars in these studies form a heterogenous sample that covers a
wide range in effective temperatures, major element composition (H or
He)\footnote{DA and DB denote white dwarfs with hydrogen and
  helium-dominated atmospheres, respectively. The letter ``Z'' is
  added to identify the presence of photospheric metals.}, and
metal settling timescales from days to millions of years. In
particular for those objects with extended convection zones and long
diffusion timescales (cool DA and the majority of DB white dwarfs), it
is usually not possible to determine whether accretion has just
started, if it is now ongoing in a quasistationary state, or if it
has already ended long ago. This uncertainty impedes accurate
measurements of both the accretion rates and chemical abundances of
the circumstellar debris \citep{Koester09}. Addressing the overall
picture and answering questions, such as ``How many and what
  kinds of white dwarf progenitors harbored a planetary system, and
  what were their compositions? Where does the reservoir of small
  bodies that are scattered reside, and how does it evolve with
  time?'', requires deep, high-resolution observations of a much
larger and unbiased sample. 

Ground-based observations of white dwarfs are primarily sensitive to
the detection of photospheric Ca via the H\&K
doublet. \citet{Zuckerman.Koester.ea03} carried out the first
systematic study, observing with the Keck HIRES echelle spectrograph
primarily cool ($T_\mathrm{eff} \la$10\,000~K) and old (cooling age
$\ga$1~Gyr) DA white dwarfs, and found traces of Ca in $\approx$25\%
of their target sample. A number of additional white dwarfs with weak
Ca\,H\&K lines were identified in the SPY survey
\citep{Napiwotzki.Christlieb.ea01, Koester.Rollenhagen.ea05*b,
  Berger.Koester.ea05}, and \citet{Koester.Wilken06} used this
combined sample of 38 DAZ to test the hypothesis of interstellar
accretion as origin of the metals. A similar, but smaller study
focusing on 30 DB white dwarfs reached comparable conclusions,
$\approx$25\% of these stars have remnants of planetary systems
\citep{zuckermanetal10-1}, with the statistics subject to the caveat
regarding stars with deep convection zones mentioned above.
Fundamental limitation of all these ground-based studies are that (1)
for a given photospheric Ca abundance, the strength of the H\&K
doublet varies by many orders of magnitude for white dwarf
temperatures ranging from 10\,000~K to 25\,000~K, with corresponding
cooling ages of a few 10~Myr to nearly 1~Gyr, and (2) Ca is only a
modest component of rocky planets, $<$2\% of the bulk Earth
\citep{mcdonough00-1}.

Space-based observations of white dwarfs are sensitive to many
additional polluting elements, but have so far focused predominantly
on hot ($>$30000~K) white dwarfs. While a substantial number of hot
white dwarfs showing photospheric metals have been detected
\citep{Shipman.Provencal.ea95, Bannister.Barstow.ea03,
  Barstow.Good.ea03, Dickinson.Barstow.ea12,
  Dickinson.Barstow.ea12*b}, the interpretation of their origin has
been complicated by the strong effects of radiative levitation in
their atmospheres, though \cite{Barstow.Barstow.ea14} argue for
external pollution.

Here, we present the results of a well defined large far-ultraviolet
spectroscopic survey of warm (17\,000~--~27\,000~K) DA white dwarfs
that is sensitive to many of the major components of rocky material
(O, Fe, and in particular Si) and extends our knowledge regarding the
frequency of planetary debris around white dwarfs and their main
sequence progenitors to younger cooling
ages ($\approx$20--200~Myr). In this temperature range, DA white dwarfs
have very simple atmospheres, no convection zones, and short
diffusion timescales. It is therefore safe to assume that the
diffusion occurs in a steady state and the calculation of the chemical
composition of the accreted matter is straightforward. One
complication in this range~--~radiative levitation of silicon and
carbon~--~was not anticipated at the beginning of the project, but we
developed the necessary calculations to account for this effect.

\begin{figure*}
\centering 
\includegraphics[angle=-90,width=0.9\textwidth]{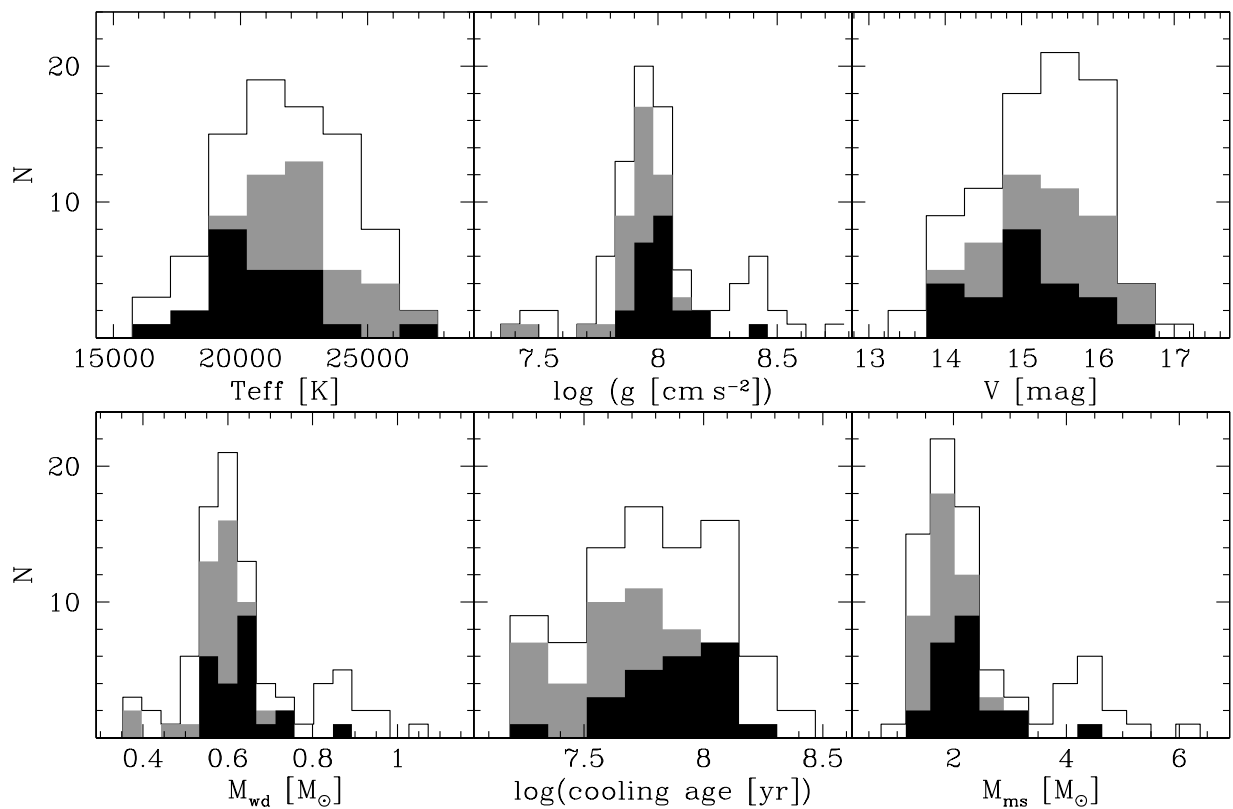}
\caption{Fundamental properties of the DA stars observed in this
  HST/COS far-ultraviolet spectroscopic survey. From top left to
  bottom right: effective temperature and surface gravity
  (Sect.~\ref{s-atmos}), $V$-band magnitude, white dwarf mass and
  cooling age computed from $T_\mathrm{ eff}$ and $\log g$ using the
  cooling models of \citet{Fontaine.Brassard.ea01}, and the
  main-sequence progenitor mass computed from $M_\mathrm{wd}$ and the
  initial-to-final mass relation of \citet{kaliraietal08-1}. The full
  sample is indicated by the outlined histogram.  48 white dwarfs
  where at least photospheric Si has been detected are shown with the
  filled histograms, of which 23 must be currently accreting
  (black). The remaining 25 (grey) have Si abundances consistent with
  support from radiative levitation alone, and have very likely
  accreted in the recent past (Sect.~\ref{s-ismaccretion}).}
\label{fig1}
\end{figure*}

\section{Sample selection and HST observations}
We selected an input target list of 150 DA white dwarfs from
\citet{Liebert.Bergeron.ea05} and \citet{Koester.Voss.ea09}, where the sole
criteria were that the stars (a) had 17\,000~K $<T_\mathrm{eff}<25\,000$~K
(two targets turned out to be slightly hotter in our analysis, extending the
temperature range to $\approx$27\,000~K), and (b) had predicted fluxes
$F_\lambda$(1300\,\AA) $ > 5\times10^{-14}~\mathrm{erg~cm}^{-2}~
\mathrm{s}^{-1}$\,\AA$^{-1}$.  Given the need to obtain far-ultraviolet
spectroscopy for a large sample of DA white dwarfs, but with no specific need
to observe any particular star, this project was implemented as an HST
Snapshot Program. Such snapshots are short observations occupying at most one
HST orbit, which are executed in gaps of the scheduled observing sequence,
where no other target from the pool of accepted regular GO or Large programs
is available. We included several types of ancillary targets to facilitate the
comparison with the main DA sample, and to test our computational tools:
several white dwarfs with close M-dwarf companions (drawn from
\citealt{schreiber+gaensicke03-1} and \citealt{farihietal10-3}), two recently
discovered metal-polluted DA white dwarfs from \citet{Vennes.Kawka.ea10,
  vennesetal11-1}, and a small number of bright DB white dwarfs from
\citet{bergeronetal11-1}. This HST program was executed from September 2010 to
February 2013, and a total of 85 DA white dwarfs from our target list, plus 15
ancillary targets were successfully observed.  The distribution of the
fundamental parameters of the 85 DA white dwarfs ($T_\mathrm{eff}$, $\log g$,
$V$-band magnitude, $M_\mathrm{wd}$, cooling age, and progenitor mass
$M_\mathrm{ms}$) are illustrated in Fig.~\ref{fig1}.

All observations were obtained with the Cosmic Origin Spectrograph
(COS, \citealt{greenetal12-1}) using the G130M grating. We adopted a
central wavelength of 1291\,\AA, which covers the wavelength range
from 1130 to 1435\,\AA, with a gap at 1278$-$1288\,\AA\ due to the
space between the two detector segments. The COS detector suffers of a
number of fixed-pattern problems, which can be eliminated by obtaining
multiple spectra dithered by small steps in the dispersion direction,
so-called FP-POS settings. The best results are obtained using all
available four FP-POS settings, but given the limited time available in
a snapshot exposure, we chose to split the exposure time equally
between only two FP-POS positions (1 \& 4).

Several of our COS observations were already used for the study of
individual objects, including a detailed abundance study of four
strongly metal-polluted DAZ stars that also exhibit infrared excess from
circumstellar dust \citep{Gansicke.Koester.ea12}; the detection of
planetary debris at two white dwarfs in the Hyades
(\citealt{Farihi.Gansicke.ea13}, see also Sect.~\ref{s-hyadeswd}) and
of water in an extrasolar minor planet \citep{farihietal13-2}; and the
identification of molecular hydrogen in one DA white dwarf that turned
out to be significantly cooler than the published effective
temperatures \citep{xuetal13-2, zuckermanetal13-1}.

In this paper, we focus on the statistical analysis of the DA sample,
a more detailed discussion of the white dwarfs that exhibit
photospheric absorption in addition to Si and C will be discussed
elsewhere. 

\section{Atmospheric parameters\label{s-atmos}}
The parameters were determined by comparison of the observed spectra
with a grid of theoretical models, calculated with the methods and
input physics as described in \cite{Koester10}. The only important
change compared to the grid used for the analysis of the SPY survey
\citep{Koester.Voss.ea09} are the inclusion of the non-ideal Balmer
line Stark profiles as calculated by \cite{Tremblay.Bergeron09} and
new Lyman $\alpha$ profiles.

The new calculations for the broadening of the Lyman $\alpha$ line,
and in particular the quasi-molecular satellites at 1400 and 1600\,\AA,
use the unified theory as described in \cite{Allard.Kielkopf82} and
\cite{Allard.Royer.ea99}. The program code for the profiles, however,
has been completely rewritten with improved numerical algorithms to
overcome problems with numerical noise and artifacts, caused by the
very large dynamical range of ten or more orders of magnitude between
the line core absorption and the far wing in very broad lines. We have
also used new calculations for the adiabatic potentials and dipole
moments for the H-H+ interaction by \cite{Santos.Kepler12}.

The HST/COS spectra are dominated by the Lyman~$\alpha$ profile,
including the satellite near 1400\,\AA, and contain practically no
undisturbed continuum. Although a fit can be obtained with effective
temperature $T_\mathrm{eff}$\ and surface gravity $\log g$\ as free
parameters, the result is quite uncertain. A small increase in $\log
g$\ can almost perfectly be compensated by an increase in
$T_\mathrm{eff}$, such that the ionization fraction, the major
determinant for atmospheric structure and line strengths, remains
constant. Very minor changes in the fitting routine cause the solution
to wander along this correlation line. It is therefore necessary to
use additional information to fix either the temperature or $\log
g$\ in the fitting of the COS spectra. Fortunately all objects have
recent parameter determinations from high quality optical spectra,
which do not suffer from this degeneracy, and up-to-date model
atmospheres.

Seventy one of the 90 objects (including five of the ancillary
targets) have been analyzed from optical spectra by
\cite{Gianninas.Bergeron.ea11} (hereafter G11), and 60 objects have at least
two spectra in the SPY database. The latter have been reanalyzed by us
using the latest model grid. The overlap of both samples (excluding
WD0933+025 and WD1049+103 with strong contribution from a red
companion in the optical [G11]) contains 39 objects, which
can be used to estimate systematic and statistical errors of the
parameter determination.

\begin{figure}
\centering
\includegraphics[width=0.45\textwidth]{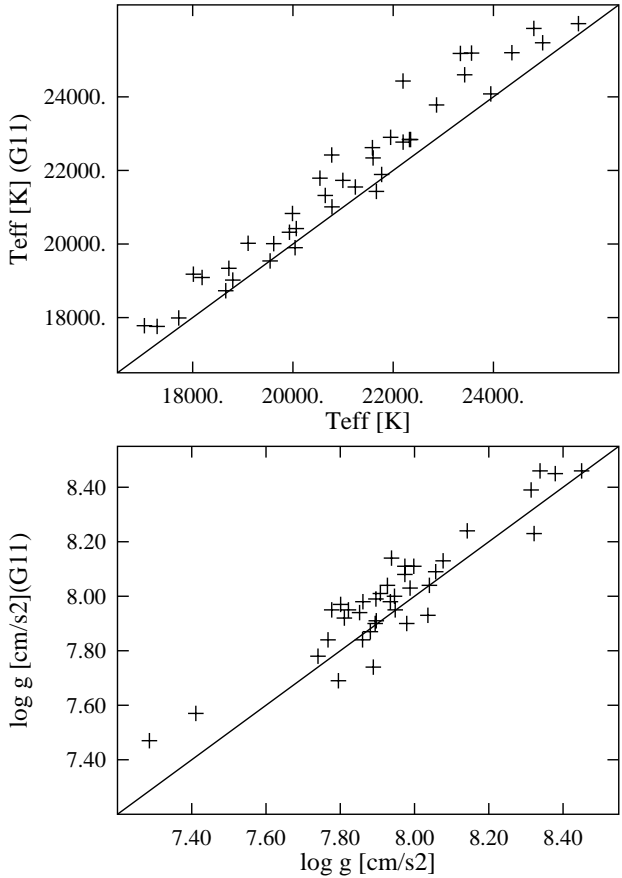}
\caption{Comparison of $T_\mathrm{eff}$\ and $\log g$\ between the
  \cite{Gianninas.Bergeron.ea11} results and our own determinations
  from the reanalysis of the SPY spectra. The continuous lines are the
  1:1 relations.} \label{fig2}
\end{figure}

The G11 temperatures are on average higher by 3.16\%, the surface gravities
are larger by 0.056~dex. One possible explanation for these differences could
be the theoretical models, but we believe that we use essentially the same
input physics as the Montreal group. The models are rather simple with no
convection, fairly low matter densities, and weak non-ideal effects. When we
implemented the \cite{Tremblay.Bergeron09} Stark broadening data, P. Tremblay
(2009, private communication) provided a theoretical spectrum for the optical
range of a DA at 18\,000~K, $\log g$\ = 8. The corresponding model in our
current grid is within $<1\%$ of the flux identical to that benchmark
spectrum. We thus do not believe that the different results come from
differences in the theoretical models. The different nature of the
observations (e.g. echelle vs. long slit spectra), or different fitting
methods remain as possible explanations.

Apart from the systematic differences there is also a statistical
scatter, which we interpret as an indication of the statistical error
in our parameter determination ($\sigma(T_\mathrm{eff}) \approx$2.36\%,
$\sigma(\log g) \approx$0.084). Fig.~\ref{fig2} shows the
comparison for the common sample.

The systematic difference in the optical parameter determination between
  G11 and the current work leads to different predictions for the theoretical
  UV spectra. The $T_\mathrm{eff}$ difference has a much larger effect than
  that of $\log g$; we have therefore decided to adopt the surface gravities
from optical results and use $T_\mathrm{eff}$\ as free fit parameter for the
COS spectra. Keeping our own optical surface gravities fixed we obtain
temperatures from the ultraviolet fitting, which are similar to our optical
temperatures, with an average difference of only 80~K. On the other hand,
using the G11 surface gravities, our fits to the COS spectra results in
temperatures, which are on average 600~K lower than found by G11 from their
optical spectra. In simple words: using our own models and the described
method, the parameters are consistent between optical and ultraviolet fitting,
whereas for the G11 parameters they are not.  We do not know if a fit of the
ultraviolet spectra with the Montreal models would give more consistent
results, but the implication would be that their theoretical Lyman $\alpha$
spectra are different from our current implementation.

The derived (COS) temperatures using either the G11 or our $\log g$\ 
values from the SPY spectra differ by only 169~K on average, much
less than they differ from the G11 temperatures, as shown above, with
a distribution completely determined by the statistical errors of the
optical surface gravities. The statistical (formal) errors for
$T_\mathrm{eff}$\ from the fit routines are small (about 20~K
typically), and can be neglected entirely. This results in a tight
correlation between $\log g$ and the resulting COS temperature, which we
use in the error estimates below.

\begin{figure}
\centering
\includegraphics[width=0.45\textwidth]{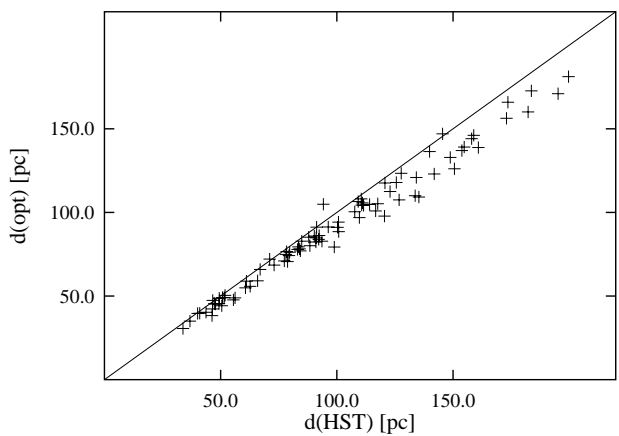}
\caption{Comparison of distances derived using the $V$ magnitude
  vs. using the absolute calibration of COS. The systematic deviation
  from the 1:1 relation (solid line) is caused by our neglect of
  interstellar reddening with a much stronger effect in the ultraviolet than
  on $V$.} \label{fig3}
\end{figure}

These experiments suggested the following procedure. For the
overlapping sample between G11 and SPY we take the average $\log
g$. For those with only G11 determination we decrease $\log g$\ by one
half the systematic difference (0.028), whereas for those with SPY
determination alone, we increase $\log g$\ by the same amount. This
decreases the systematic difference to either sample to $\pm 0.028$ in
$\log g$; adding the statistical error quadratically results in
$\sigma$($\log g$) = 0.089. With this $\log g$\ kept fixed, we fit the
ultraviolet spectra to determine a (COS) temperature. The error of $\log
g$\ translates into an error of $T_\mathrm{eff}$\ of 1.36\%. We
therefore use as our final error estimates $\sigma(T_\mathrm{eff}) =
1.36\%$ and $\sigma(\log g) = 0.089$, with a very tight correlation
(larger $\log g$ leads to larger $T_\mathrm{eff}$).  The results are
presented in Table~\ref{parameters}, which also contains photometric
$V$ data from the SIMBAD database, the Villanova White Dwarf
Catalog\footnote{\url{http://www.astronomy.villanova.edu/WDcatalog/}},
and the APASS Photometric All-Sky
Survey\footnote{\url{http://www.aavso.org/apass}},  
and some derived quantities. Masses
and radii are obtained from the Montreal evolutionary
calculations\footnote{\url{http://www.astro.umontreal.ca/~bergeron/CoolingModels/}}  
\citep{Fontaine.Brassard.ea01}, two independent distance estimates
from the comparison of the absolute HST flux and the $V$ magnitude
with theoretical predictions from the model atmosphere parameters and
the radius.  A comparison of both values is shown in
Fig.~\ref{fig3}. 

The systematic difference between the two values indicates that some
  amount of reddening is present for the more distant objects. Assuming a
  $1/\lambda$ dependence leads to an $\approx$20\% change of the absorption
  over the broad L$\alpha$ line. In our spectral fitting we use the fixed
  surface gravity from the optical determination. The temperature is then
  obtained from a fit to the L$\alpha$ line, which is dominated by the overall
  strength (i.e. the equivalent width) of this line, unaffected by the
  reddening.

\begin{longtab}
\begin{landscape}
\begin{longtable}{lrrrrrrrl}
\caption{General data for the white dwarfs sample. $T_\mathrm{eff}$
  and $\log g$\ come from a comparison of optical and HST/COS spectra
  with theoretical models. Mass and radius are derived from these
  using the Montreal evolutionary calculations and finite temperature
  mass-radius relations. The uncertainties (data in parentheses) are
  propagated and include possible systematic as well as statistical
  uncertainties.  Distances are derived from the absolute flux in the
  ultraviolet (dHST) and from observed V (dV) and theoretical
  models. Uncertainties again include systematic and statistical
  contributions as far as possible. Below the horizontal line are the
  five comparison objects.}
\label{parameters}\\ 
\hline\hline 
 object    &      $T_\mathrm{eff}$ [K]  &   $\log g$ [cm\,s$^{-2}$]
 &  $V$ [mag] &  
  $M/M_\sun$ &  100~$R/R_\sun$  & dHST [pc] & dV [pc]& Notes \\
\hline
\endfirsthead
\caption{continued.}\\
\hline\hline
 object    &      $T_\mathrm{eff}$ [K] &   $\log g$ [cm\,s$^{-2}$] &  
   $V$ [mag]&  $M/M_\sun$ &  100~$R/R_\sun$  & dHST [pc] & dV [pc]& Notes \\
\hline
\endhead
\hline
\endfoot
WD0000+171     &21268 (289) &7.927 (0.089)&15.815 (0.053)&0.587
        (0.048)& 1.379 (0.085) &122.9 (8.4) & 112.5 (7.6)\\
WD0013-241     &19032 (258) &8.040 (0.089)&15.379 (0.048)&0.642 
        (0.049)& 1.267 (0.082) & 83.1 (6.0) &  77.9 (5.4)\\  
WD0018-339     &22160 (301) &8.508 (0.089)&14.639 (0.047)&0.938 
        (0.053)& 0.893 (0.066) & 50.5 (4.0) &  44.2 (3.4)\\
WD0028-474     &17550 (238) &7.760 (0.089)&15.150 (0.029)&0.492
        (0.042)& 1.531 (0.091) & 98.9 (6.8) &  79.3 (4.9)\\
WD0047-524     &18831 (256) &7.890 (0.089)&14.228 (0.050)&0.561 
        (0.047)& 1.407 (0.085) & 51.9 (3.6) &  50.4 (3.3)\\
WD0059+257     &20491 (278) &8.002 (0.089)&15.903 (0.016)&0.626 
        (0.048)& 1.307 (0.083) &110.5 (7.8) & 108.1 (7.0)\\
WD0102+095     &25001 (340) &7.932 (0.089)&14.439 (0.020)&0.599 
        (0.047)& 1.385 (0.088) & 73.0 (5.1) &  68.5 (4.5)\\
WD0114-605     &25153 (342) &7.838 (0.089)&15.114 (0.050)&0.552 
        (0.044)& 1.482 (0.093) &107.7 (7.5) & 100.4 (6.9)\\
WD0124-257     &23424 (318) &7.815 (0.089)&16.183 (0.050)&0.536 
        (0.043)& 1.500 (0.093) &173.0 (11.9)& 156.3 (0.6)\\
WD0140-392     &23174 (315) &7.983 (0.089)&14.346 (0.025)&0.622 
        (0.047)& 1.331 (0.085) & 65.9 (4.7) &  59.1 (3.9)\\
WD0155+069     &21435 (291) &7.804 (0.089)&15.466 (0.369)&0.525 
        (0.043)& 1.503 (0.092) & 94.2 (6.5) & 104.9 (9.0)\\
HS0200+2449    &22276 (302) &7.981 (0.089)&15.713 (0.080)&0.619 
        (0.048)& 1.331 (0.085) &126.8 (8.9) & 107.5 (8.0)\\
WD0242-174     &20980 (285) &7.973 (0.089)&15.384 (0.010)&0.611 
        (0.048)& 1.335 (0.084) &100.8 (7.1) &  88.5 (5.7)\\
WD0307+149     &21976 (298) &7.983 (0.089)&15.384 (0.258)&0.619 
        (0.048)& 1.328 (0.085) & 91.3 (6.5) &  91.2 (12.3)\\
WD0308+188     &17865 (242) &7.893 (0.089)&14.191 (0.037)&0.559 
        (0.048)& 1.400 (0.084) & 46.6 (3.2) &  47.4 (3.0)\\
HE0308-2305    &25066 (340) &8.590 (0.089)&15.076 (0.058)&0.981 
        (0.044)& 0.831 (0.067) & 62.7 (5.3) &  55.8 (4.8)\\
WD0341+021     &22508 (306) &7.378 (0.089)&15.407 (0.063)&0.368 
        (0.025)& 2.056 (0.141) &158.0 (11.9)& 144.1 (10.8)\\
WD0352+018     &21647 (294) &7.863 (0.089)&15.569 (0.086)&0.555 
        (0.045)& 1.444 (0.089) &109.2 (7.6) & 106.5 (7.9)\\
HE0358-5127    &24607 (334) &8.070 (0.089)&15.717 (0.055)&0.670 
        (0.048)& 1.250 (0.084) &133.7 (9.8) & 110.0 (8.0)\\
HE0403-4129    &24000 (326) &8.051 (0.089)&16.351 (0.172)&0.659 
        (0.048)& 1.267 (0.084) &158.9 (11.5)& 146.1 (15.2)\\
HE0414-4039    &21664 (294) &8.091 (0.089)&16.133 (0.084)&0.675 
        (0.048)& 1.225 (0.082) &120.7 (9.0) & 117.6 (9.2)\\
HE0416-1034    &27168 (369) &8.039 (0.089)&15.368 (0.031)&0.660 
        (0.047)& 1.285 (0.086) &117.6 (8.5) & 105.2 (7.3)\\
HE0418-1021    &24086 (327) &8.370 (0.089)&15.679 (0.053)&0.846 
        (0.059)& 0.994 (0.067) & 91.1 (6.7) &  84.8 (6.2)\\
WD0421+162     &19322 (262) &8.104 (0.089)&14.289 (0.044)&0.677 
        (0.049)& 1.208 (0.080) & 47.3 (3.5) &  45.5 (3.2)\\
WD0431+126     &20844 (283) &8.054 (0.089)&14.226 (0.030)&0.653 
        (0.048)& 1.257 (0.082) & 49.4 (3.6) &  48.7 (3.3)\\
HE0452-3444    &21810 (296) &7.887 (0.089)&16.128 (0.050)&0.567 
        (0.046)& 1.420 (0.088) &139.9 (9.8) & 136.4 (9.1)\\
HS0507+0434A   &21529 (292) &8.080 (0.089)&14.222 (0.059)&0.669 
        (0.048)& 1.234 (0.082) & 56.2 (3.7) &  48.9 (3.6)\\
WD0843+516     &22412 (304) &7.902 (0.089)&16.044 (0.046)&0.577 
        (0.047)& 1.407 (0.087) &148.8 (10.2)& 132.9 (8.8)\\
WD0854+404     &22813 (310) &7.932 (0.089)&14.811 (0.023)&0.593 
        (0.047)& 1.379 (0.086) & 78.1 (5.5) &  74.9 (4.8)\\
WD0920+363     &24768 (336) &7.672 (0.089)&16.070 (0.050)&0.477 
        (0.038)& 1.667 (0.104) &183.7 (12.8)& 172.7 (11.7)\\
WD0933+025     &20987 (285) &7.821 (0.089)&15.928 (0.028)&0.532 
        (0.044)& 1.484 (0.091) &150.6 (10.4)& 126.1 (8.0)& 3\\
HS0944+1913    &17027 (231) &8.003 (0.089)&14.508 (0.024)&0.618 
        (0.049)& 1.297 (0.081) & 51.5 (3.7) &  49.0 (3.2)\\
WD0947+325     &22340 (303) &8.312 (0.089)&15.504 (0.099)&0.805 
        (0.057)& 1.037 (0.070) & 78.4 (5.7) &  76.6 (6.3)\\
WD0954+697     &21831 (296) &7.922 (0.089)&15.962 (0.068)&0.586 
        (0.047)& 1.386 (0.086) &127.7 (8.9) & 123.4 (8.7)\\
WD1005+642     &19450 (264) &7.922 (0.089)&13.692 (0.023)&0.579 
        (0.048)& 1.378 (0.084) & 40.1 (2.8) &  39.6 (2.5)\\
WD1013+256     &22133 (301) &8.022 (0.089)&16.316 (0.098)&0.640 
        (0.048)& 1.291 (0.084) &153.8 (11.0)& 137.0 (10.9)\\
WD1015+161     &18911 (257) &8.042 (0.089)&15.607 (0.061)&0.643 
        (0.049)& 1.264 (0.082) & 90.3 (6.5) &  85.9 (6.1)\\
WD1017+125     &21500 (292) &7.949 (0.089)&15.674 (0.046)&0.599 
        (0.048)& 1.359 (0.085) &114.1 (8.0) & 104.8 (7.0)\\
WD1034+492     &20682 (281) &8.192 (0.089)&15.433 (0.042)&0.729 
        (0.052)& 1.133 (0.075) & 79.7 (5.9) &  76.2 (5.4)\\
WD1038+633     &24295 (330) &8.392 (0.089)&15.152 (0.060)&0.861 
        (0.060)& 0.978 (0.066) & 67.0 (4.9) &  65.9 (4.9)\\
WD1049+103     &20388 (277) &7.804 (0.089)&15.808 (0.121)&0.522 
        (0.044)& 1.499 (0.091) &125.6 (8.5) & 117.9 (9.8)& 3\\
WD1049-158     &19634 (267) &8.352 (0.089)&14.357 (0.008)&0.827 
        (0.059)& 1.003 (0.067) & 41.0 (3.0) &  39.7 (2.7)\\
WD1052+273     &23789 (323) &8.392 (0.089)&14.120 (0.050)&0.860 
        (0.061)& 0.978 (0.066) & 43.7 (3.2) &  40.3 (2.9)\\
WD1058-129     &25218 (342) &8.749 (0.089)&14.912 (0.011)&1.063 
        (0.048)& 0.720 (0.058) & 47.8 (4.1) &  45.2 (3.7)\\
WD1102+748     &19777 (268) &8.352 (0.089)& 0.000 (0.000)&0.827 
        (0.059)& 1.004 (0.067) & 59.0 (4.3) &   0.0 (0.0)\\
WD1104+602     &18206 (247) &8.052 (0.089)&13.740 (0.110)&0.646 
        (0.049)& 1.253 (0.081) & 36.8 (2.7) &  35.0 (2.9)& \\
WD1129+155     &18915 (257) &8.191 (0.089)&14.091 (0.047)&0.725 
        (0.053)& 1.131 (0.075) & 46.3 (3.4) &  38.3 (2.7)\\
WD1133+293     &23671 (321) &7.862 (0.089)&14.883 (0.066)&0.560 
        (0.045)& 1.452 (0.091) & 92.2 (6.4) &  83.9 (5.9)\\
WD1229-013     &20652 (280) &7.532 (0.089)&14.460 (0.063)&0.408 
        (0.032)& 1.812 (0.114) & 84.3 (6.0) &  77.0 (5.4)\\
WD1230-308     &25501 (346) &8.276 (0.089)&15.726 (0.031)&0.789 
        (0.055)& 1.070 (0.072) &120.5 (8.8) &  97.8 (6.9)\\
HS1243+0132    &21871 (297) &7.881 (0.089)&16.463 (0.203)&0.564 
        (0.046)& 1.426 (0.088) &182.3 (12.7)& 160.1 (18.0)\\
WD1257+048     &22727 (309) &7.982 (0.089)&15.020 (0.050)&0.620 
        (0.048)& 1.331 (0.085) & 83.5 (6.0) &  79.4 (5.5)\\
WD1310-305     &20021 (272) &7.903 (0.089)&14.478 (0.023)&0.570 
        (0.048)& 1.398 (0.085) & 61.1 (4.1) &  58.9 (3.7)\\
WD1325+279     &20918 (284) &7.942 (0.089)&15.800 (0.050)&0.594  
        (0.048)& 1.364 (0.085) &135.3 (9.5) & 109.2 (7.3)\\
WD1325-089     &17133 (233) &7.903 (0.089)& 0.000 (0.000)&0.563 
        (0.048)& 1.388 (0.083) & 67.6 (4.7) &   0.0 (0.0)\\
WD1330+473     &22741 (309) &7.922 (0.089)&15.233 (0.025)&0.588 
        (0.047)& 1.389 (0.087) & 96.3 (6.8) &  91.3 (5.9)\\
WD1353+409     &24133 (328) &7.562 (0.089)&15.487 (0.017)&0.431 
        (0.034)& 1.800 (0.114) &154.8 (10.8)& 139.1 (9.1)\\
WD1408+323     &18784 (255) &7.972 (0.089)&13.970 (0.050)&0.605 
        (0.048)& 1.330 (0.083) & 46.5 (3.3) &  42.3 (2.8)\\
WD1451+006     &26618 (362) &7.940 (0.089)&15.266 (0.050)&0.607 
        (0.046)& 1.382 (0.089) &111.0 (7.7) & 105.7 (7.4)\\
WD1459+347     &21671 (294) &8.502 (0.089)&15.794 (0.015)&0.934 
        (0.054)& 0.898 (0.066) & 79.4 (6.4) &  74.2 (5.5)\\
WD1524-749     &23547 (320) &7.743 (0.089)&15.930 (0.050)&0.504 
        (0.041)& 1.579 (0.098) &145.5 (10.1)& 147.0 (9.9)\\
WD1531-022     &19472 (264) &8.455 (0.089)&13.973 (0.034)&0.898 
        (0.058)& 0.929 (0.065) & 33.8 (2.6) &  30.6 (2.2)\\
WD1547+057     &24566 (334) &8.414 (0.089)&15.940 (0.053)&0.876 
        (0.061)& 0.962 (0.065) &100.7 (7.4) &  94.2 (6.9)\\
WD1548+149     &21686 (294) &7.897 (0.089)&15.156 (0.029)&0.572 
        (0.047)& 1.410 (0.087) & 92.4 (6.5) &  86.2 (5.5)\\
WD1633+676     &23685 (322) &8.042 (0.089)&16.250 (0.050)&0.654 
        (0.048)& 1.275 (0.084) &160.9 (11.8)& 138.8 (9.8)\\
WD1647+375     &22803 (310) &7.902 (0.089)&14.913 (0.038)&0.578 
        (0.047)& 1.408 (0.087) & 88.4 (6.2) &  80.1 (5.3)\\
WD1713+332     &21873 (297) &7.462 (0.089)&14.387 (0.092)&0.389 
        (0.028)& 1.919 (0.127) & 91.0 (6.7) &  82.3 (6.6)\\
WD1755+194     &25005 (340) &7.886 (0.089)& 0.000 (0.000)&0.575 
        (0.045)& 1.432 (0.090) & 67.6 (11.7)&   0.0 (0.0)\\
WD1914-598     &19231 (261) &7.850 (0.089)& 0.000 (0.000)&0.541 
        (0.046)& 1.447 (0.088) & 57.1 (3.9) &   0.0 (0.0)\\
WD1943+163     &19451 (264) &7.896 (0.089)&13.951 (0.027)&0.565
        (0.047)& 1.403 (0.085) & 49.3 (3.4) &  45.4 (2.9)\\
WD1953-715     &18975 (258) &7.957 (0.089)&15.057 (0.062)&0.597 
        (0.048)& 1.344 (0.083) & 77.4 (5.4) &  71.0 (4.9)\\
WD2021-128     &20892 (284) &7.943 (0.089)&15.204 (0.078)&0.594 
        (0.048)& 1.363 (0.085) & 84.9 (6.0) &  82.8 (6.0)\\
WD2032+188     &17878 (243) &7.490 (0.089)&15.318 (0.025)&0.383
        (0.030)& 1.843 (0.117) &111.5 (8.0) & 104.2 (6.8)\\
WD2046-220     &24362 (331) &7.959 (0.089)&15.374 (0.049)&0.612 
        (0.047)& 1.357 (0.087) &116.7 (8.2) & 100.9 (7.0)\\
WD2058+181     &17308 (235) &7.920 (0.089)&15.198 (0.063)&0.572 
        (0.049)& 1.373 (0.082) & 71.1 (4.9) &  72.1 (4.9)\\
WD2134+218     &17970 (244) &8.039 (0.089)&14.477 (0.094)&0.639 
        (0.049)& 1.265 (0.081) & 50.8 (3.7) &  49.1 (3.8)\\
HS2210+2323    &23967 (325) &8.390 (0.089)&15.876 (0.081)&0.859 
        (0.060)& 0.979 (0.066) &100.3 (7.4) &  91.1 (7.1)\\
WD2220+133     &23301 (316) &8.399 (0.089)&15.612 (0.050)&0.864 
        (0.061)& 0.972 (0.065) & 84.0 (6.1) &  78.2 (5.6)\\
HS2229+2335    &18820 (255) &8.023 (0.089)&16.008 (0.066)&0.633 
        (0.049)& 1.282 (0.082) &110.7 (7.9) & 104.4 (7.5)\\
HE2231-2647    &21126 (287) &7.796 (0.089)&14.966 (0.044)&0.521 
        (0.043)& 1.510 (0.092) & 93.6 (6.4) &  82.8 (5.4)\\
HE2238-0433    &16756 (227) &8.233 (0.089)&16.897 (0.025)&0.746 
        (0.055)& 1.094 (0.072) &141.9 (10.6)& 123.0 (8.4)\\
HS2244+2103    &23995 (326) &7.937 (0.089)&16.449 (0.125)&0.599 
        (0.047)& 1.378 (0.087) &173.6 (12.2)& 165.9 (14.3)\\
WD2306+124     &20645 (280) &8.073 (0.089)&15.078 (0.049)&0.663 
        (0.048)& 1.239 (0.082) & 78.8 (5.8) &  70.6 (5.0)\\
WD2322-181     &21862 (297) &8.009 (0.089)&15.284 (0.047)&0.633 
        (0.048)& 1.303 (0.084) & 87.9 (6.3) &  85.2 (5.9)\\
WD2359-324     &24879 (338) &7.890 (0.089)&16.410 (0.050)&0.577 
        (0.046)& 1.427 (0.090) &195.2 (13.6)& 174.1 (11.9)\\
\hline
SDSS1228+1040  &20713 (281) &8.150 (0.089)& 0.000 (0.000)&0.705 
        (0.051)& 1.169 (0.078) &134.2 (9.9) & 120.9 (9.4) * \\
WD1929+012     &21457 (291) &7.900 (0.089)&14.200 (0.050)&0.573 
        (0.047)& 1.406 (0.086) & 60.7 (4.2) &  54.9 (3.7)& *,1\\
WD0710+741     &19749 (268) &7.492 (0.089)&14.979 (0.062)&0.390  
        (0.030)& 1.856 (0.120) &109.7 (7.9) &  96.9 (6.9)& *,2\\
WD2256+249     &22500 (306) &7.812 (0.089)&13.680 (0.050)&0.532 
        (0.043)& 1.500 (0.093) & 55.5 (3.8) &  47.7 (3.2)& *,2\\
WD2257+162     &24287 (330) &7.512 (0.089)&15.973 (0.058)&0.413 
        (0.031)& 1.866 (0.122) &199.7 (14.3)& 181.2 (13.0)& *,2\\

\end{longtable}
\tablefoot{\\ *: ancillary target, not used in sample statistics; 1:
  $\log g$ from \cite{Vennes.Kawka.ea11}; 2: post-common envelope
  (close) binary; 3: spatially resolved (wide) binary; 4: exhibits
  excess infrared emission}
\end{landscape}
\end{longtab}

\begin{figure}
\centering
\includegraphics[width=0.45\textwidth]{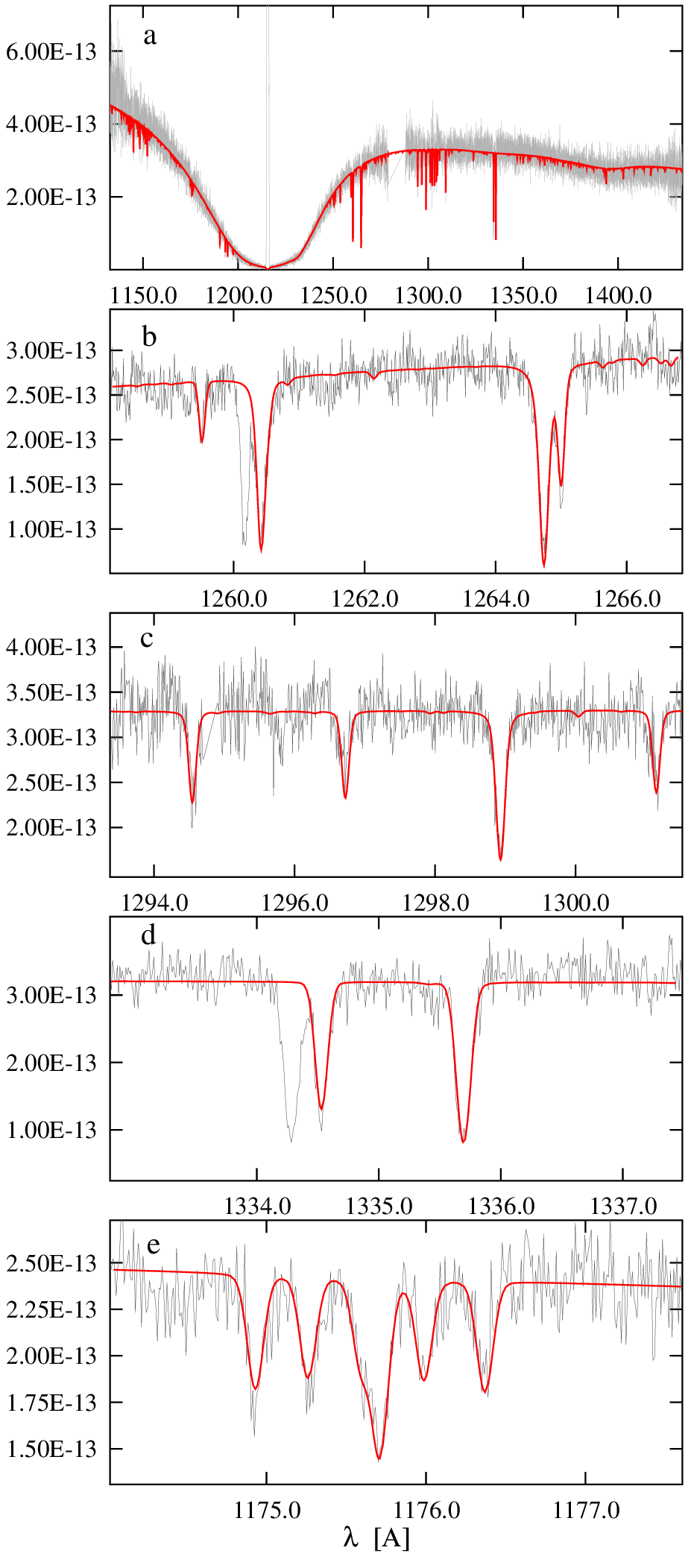}
\caption{Example spectra with the important lines for this
  work. The vertical axis is the observed flux in
  erg\,cm$^{-2}$\,s$^{-1}$\,\AA$^{-1}$.}\label{fig4}
\end{figure}

\section{Element abundances}
Atmospheric structures were calculated for the best fit parameters
with the same input physics used in constructing the
grid. With this model structure synthetic spectra were calculated
including approximately 14\,000 spectral lines of 14 elements.  The
elements were assumed to be homogeneously distributed throughout the
atmosphere. Individual abundances were then varied until a
satisfactory fit to the spectral lines in the observed spectrum was
achieved. We then calculated models with the metal abundances changed
by $\pm0.2$~dex and estimated the abundance errors from these models.
Table~\ref{abund} shows the results for the Si and C abundances.
Fig.~\ref{fig4} shows an example for the total range of the
observed spectra, as well as details of the spectral ranges with the
Si and C lines important for the present study, compared with model
fits. The panels from top to bottom show:\\
a) complete COS spectral range for WD\,1943+163 with typical
  signal-to-noise and relatively large Si and C abundances. \\ 
b) \ion{Si}{ii}\,1260.422, 1264.738, 1265.002\,\AA. Note the
  strong blue-shifted interstellar component of the resonance line
  1260.422\,\AA. The sulfur \ion{S}{i} 1259.519\,\AA\ line is also
  visible at the photospheric position. \\
c) \ion{Si}{iii}~1294.545, 1296.726, 1298.892, 1298.946,
  1301.149\,\AA\ lines. \\ 
d) carbon, \ion{C}{ii}~1334.530,
  1335.660, 1335.708\,\AA\ lines, with interstellar component of the
  resonance line 1334.530\,\AA. \\ 
e) \ion{C}{iii}~1174.930,
  1175.260, 1175.590, 1175.710, 1175.987, 1176.370\,\AA\ lines. These
  \ion{C}{iii} lines are weak in WD\,1943+163; the spectrum shown is
  for HE\,0416-1034. The continuous (red) lines are the theoretical
  model. 

\begin{table*}
\caption{Si and C abundances. Abundances are logarithm of the metal to
  hydrogen ratio by numbers. Numbers in parenthesis are estimated
  errors. If absent, the abundances are upper limits
  (Sect.~\ref{discussion}).}
 \label{abund} 
\small \centering
\begin{tabular}{llll|llll}
\hline\hline
\noalign{\smallskip}
 object & $\log$[Si/H]&$\log$[C/H]& Notes & object &
                 $\log$[Si/H]&$\log$[C/H]& Notes\\
\hline
\noalign{\smallskip}
WD0000+171   & -8.10 ~(0.15) & -8.30        & &
WD1102+748   & -8.50       &                &\\
WD0013-241   & -8.40       &                & &
WD1104+602   & -8.40       &                & \\
WD0018-339   & -8.00       &                & &
WD1129+155   & -8.30 ~(0.25) & -8.10        &\\
WD0028-474   & -8.40       &                & &
WD1133+293   & -7.70 ~(0.15) & -7.70        &\\
WD0047-524   & -8.40       &                & &
SDSS1228+1040& -5.20~(0.10)& -7.60~(0.20)   & *, 1, 4, 5\\
WD0059+257   & -6.50 ~(0.10) & -8.00        & 1&
WD1229-013   & -8.30       &                &\\
WD0102+095   & -7.35 ~(0.15) & -8.00 ~(0.20)& &
WD1230-308   & -8.10       &                &\\
WD0114-605   & -7.40 ~(0.20) & -8.10 ~(0.20)& &
HS1243+0132  & -7.80 ~(0.15) & -8.30        &\\
WD0124-257   & -8.20       &                & &
WD1257+048   & -7.50 ~(0.15) & -7.70 ~(0.20)& \\
WD0140-392   & -7.50 ~(0.15) & -8.40        & &
WD1310-305   & -7.70 ~(0.15) & -8.10        &\\
WD0155+069   & -8.30       &                & &
WD1325+279   & -8.00 ~(0.15) & -8.20        &\\
HS0200+2449  & -8.30       &                & &
WD1325-089   & -8.60 ~(0.25) & -8.00        &\\
WD0242-174   & -8.20 ~(0.20) & -8.30        & &
WD1330+473   & -8.30       &                &\\
WD0307+149   & -7.85 ~(0.15) & -8.30        & &
WD1353+409   & -7.80       &                &\\
WD0308+188   & -8.20       &                & &
WD1408+323   & -7.70 ~(0.15) & -8.30 ~(0.20)& \\
HE0308-2305  & -7.80       &                & &
WD1451+006   & -7.70 ~(0.20) & -8.20        &\\
WD0341+021   & -7.95 ~(0.20) & -7.55 ~(0.15)& &
WD1459+347   & -8.20       &                &\\
WD0352+018   & -8.10 ~(0.15) & -8.30        & &
WD1524-749   & -7.90       &                &\\
HE0358-5127  & -7.80       &                & &
WD1531-022   & -8.40       &                &\\
HE0403-4129  & -7.10 ~(0.15) & -7.90        & &
WD1547+057   & -8.20       &                &\\
HE0414-4039  & -8.20       &                & &
WD1548+149   & -7.65 ~(0.15) & -7.40 ~(0.15)& \\
HE0416-1034  & -7.35 ~(0.15) & -7.05 ~(0.20)& &
WD1633+676   & -7.90       &                &\\
HE0418-1021  & -8.20       &                & &
WD1647+375   & -6.20 ~(0.10) & -5.50 ~(0.20)& 1 \\
WD0421+162   & -7.50 ~(0.15) & -8.30        & &
WD1713+332   & -8.30 ~(0.25) & -7.60 ~(0.20)& \\
WD0431+126   & -8.10 ~(0.15) & -8.30        & &
WD1755+194   & -7.30 ~(0.20) & -8.10 ~(0.20)& \\
HE0452-3444  & -7.60 ~(0.15) & -8.20        & &
WD1914-598   & -8.30 ~(0.15) & -8.30        &\\
HS0507+0434A & -8.40 ~(0.15) & -8.50        & &
WD1929+012   & -4.75 ~(0.15) & -6.80 ~(0.30)& *, 1, 4\\
WD0710+741   & -6.70 ~(0.15) & -5.90 ~(0.30)& *, 1, 2&
WD1943+163   & -6.80 ~(0.10) & -6.60 ~(0.20)& 1\\
WD0843+516   & -5.20 ~(0.10) & -7.20 ~(0.20)& 1, 4&
WD1953-715   & -6.70 ~(0.10) & -6.60 ~(0.20)& 1\\
WD0854+404   & -7.90 ~(0.20) & -8.10        & &
WD2021-128   & -8.30       &                &\\
WD0920+363   & -7.30 ~(0.20) & -7.70        & &
WD2032+188   & -8.20       &                &\\
WD0933+025   & -8.30       &                & 3&
WD2046-220   & -7.60 ~(0.15) & -7.60        &\\
HS0944+1913  & -8.40       &                & &
WD2058+181   & -7.50 ~(0.10) & -6.50 ~(0.20)& 1\\
WD0947+325   & -8.30       &                & &
WD2134+218   & -8.10 ~(0.15) & -8.10        &\\
WD0954+697   & -8.00 ~(0.20) & -8.20        & &
HS2210+2323  & -8.20       &                &\\
WD1005+642   & -8.20       &                & &
WD2220+133   & -8.20       &                &\\
WD1013+256   & -6.90 ~(0.10) & -6.40 ~(0.20)& &
HS2229+2335  & -6.50 ~(0.15) & -6.70 ~(0.20)& \\
WD1015+161   & -6.40 ~(0.10) & -8.00        & 1, 4&
HE2231-2647  & -8.10 ~(0.15) & -8.40        &\\
WD1017+125   & -7.50 ~(0.15) & -7.60        & &
HE2238-0433  & -7.50       &                &\\
WD1034+492   & -7.55 ~(0.10) & -8.00        & &
HS2244+2103  & -7.70 ~(0.15) & -8.10        &\\
WD1038+633   & -7.90 ~(0.30) & -8.00        & &
WD2256+249   & -6.85 ~(0.10) & -6.10 ~(0.20)& *, 1, 2\\
WD1049+103   & -8.40         &              & 3 &
WD2257+162   & -7.40 ~(0.15) & -6.40 ~(0.20)& *, 2 \\
WD1049-158   & -8.40       &                & &
WD2306+124   & -7.60 ~(0.10) & -8.00 ~(0.20)& \\
WD1052+273   & -8.50       &                & &
WD2322-181   & -7.70 ~(0.10) & -7.70 ~(0.15)& \\
WD1058-129   & -8.30       &                & &
WD2359-324   & -7.80       &                &\\
\hline
\end{tabular}
\tablefoot{ *: ancillary target, not used in sample statistics; 1:
  additional elements detected among the following: O, Mg, Al, P, S,
  Ca, Fe in the ultraviolet or optical spectra; 2: post-common
  envelope (close) binary; 3: spatially resolved (wide) binary; 4:
  exhibits excess infrared emission;
  5: SDSS1228 is a gas-disc DAZ with emission lines seen in the
  infrared \ion{Ca}{ii} triplet near 8500\AA\ \citep{Gansicke.Marsh.ea06} and
  the \ion{Mg}{ii} doublet at $\approx2800$\AA\ \citep{Hartmann.Nagel.ea11}.   
 The disc emission lines are broad -- in contrast, the photospheric
absorption lines are narrow, and we would expect to detect
any emission in Si in the high-quality COS spectra of SDSS\,1228+1040.
In addition, the computations of \cite{Hartmann.Nagel.ea11}
predict little emission in \ion{Si}{ii}, and none in \ion{Si}{iii}. We use
many  \ion{Si}{ii} and \ion{Si}{iii} in the abundance analysis and do not see
a significant difference.
}
\end{table*}

Somewhat surprisingly, of the 85 sample objects in Table~\ref{abund},
48 show photospheric Si. In fact, Si is always present, if any
heavy metals are seen. 18 of these show in addition C, and 7 further
metals out of this list: Mg, Al, P, S, Ca, Cr, Fe, Ni. For reasons,
which will become clear in the following, we will concentrate in this
paper on the discussion of the Si and C abundances. The very
interesting cases with more heavy metals than these two will be
treated in separate papers.

\subsection{Effect of parameter errors on the abundances}
As a test case we have used WD\,1943+163, near the middle of the
temperature range and with Si, C, and O detected. Calculating two new
models with the parameters increased, respectively decreased by the above
uncertainties, the changes in the derived abundances are smaller than
0.04~dex for all lines, which we take as an estimate for the
uncertainty caused by parameter errors, which should be added
quadratically to the errors in Table~\ref{abund}.

\section{Interstellar absorption lines}
With the exception of SDSS\,1228+1040 and HS\,2229+2335, where the
photospheric Si lines are exceptionally strong and probably mask the
IS lines, all objects show interstellar lines of
\ion{Si}{ii}~1260\,\AA\ (and others) and \ion{C}{ii}~1334/35\,\AA. In
most objects lines from the ground state of O, N, S, Fe are also
visible. The most obvious indication for their IS nature is the absence
of \ion{Si}{ii}~1265\,\AA, which is always stronger than the
1260\,\AA\ line in the photosphere, because of the $gf$ ratio of
$\approx$1.8 and level energy only 0.036~eV above the ground state. If
\ion{Si}{ii}~1265\,\AA\ is visible, it is always photospheric only and
in almost all cases the interstellar component of
\ion{Si}{ii}~1260\,\AA\ is shifted by 0.1 to 0.3\,\AA. In addition
several stars show additional Si lines from excited states, which must
have a photospheric origin.

The situation is more complicated for the \ion{C}{ii}~1334/35\,\AA\ 
doublet, where the second level is only 0.008~eV above zero, and both
components can be observed in the ISM. Determination of photospheric
abundances is only possible when photospheric and ISM lines are
clearly separated, or if lines from excited states are observed, such
as the \ion{C}{iii}~1175\,\AA\ multiplet.

The spectral resolution of $\approx$17~km\,s$^{-1}$ does not allow 
individual components in the IS lines to be identified. However, the
equivalent widths of the IS \ion{Si}{ii}~1260\,\AA\ lines in those
objects without photospheric metals range from 21 to 203~m\AA,
corresponding to column densities of $10^{12}$ to $10^{13}$ cm$^{-2}$,
if interpreted as single line on the linear part of the curve of
growth \cite[e.g.][]{Savage.Sembach96}. These column densities are
typical values for the local interstellar cloud (LIC) surrounding our
solar system \citep{Redfield.Linsky04}; given the proximity of our
targets, it is likely the LIC or at most one or two similar small clouds
are the only absorbing ISM material between us and most of our
sample. The IS lines in these 90 objects close to the sun constitute
an important resource for the study of the ISM in our immediate
neighbourhood and will be analyzed in another paper.

\section{Interpretation of abundances in the presence of diffusion}
It is well known that the abundances in the photosphere cannot
directly be taken to infer the abundances in the accreted matter. They
are modified by the diffusion processes \citep{Koester09}. In the
temperature range of our sample (17\,000--27\,000~K) there is no
convection zone in a DA, and thus no homogeneous reservoir, which can
be used to define a diffusion timescale at the bottom of this
zone. The diffusion timescale is ill-defined in this case and depends
on the layer in the atmosphere taken as the reference~--~most
reasonably at the Rosseland optical depth 2/3. However defined in
detail, they are extremely short, of the order of days or a few years
at most. It is therefore reasonable to assume a steady state between
accretion and diffusion, which eliminates some of the uncertainties
plaguing such studies for cooler DA or DB stars with deep convection
zones. Steady state means a constant diffusion flow of each element
throughout the atmosphere, which is equal to the accretion flow
\begin{equation}
    X\, \rho\, v_\mathrm{diff} = \mathrm{const} = X_\mathrm{acc}\, 
\dot{m}_\mathrm{acc}.
\end{equation}

Here, $X$ is the mass fraction of the element in question, $\rho$ the
mass density in the atmosphere, and $v_\mathrm{diff}$ the diffusion
velocity. Since the structure of the atmosphere model is known, the
depth dependence of $\rho$ and $v_\mathrm{diff}$ can be calculated and
leads to a predicted stratified element abundance
$X(r)$. $X_\mathrm{acc}$ is the abundance in the accreted matter and
$\dot{m}_\mathrm{acc}$ the accretion flow in units of
$\mathrm{g~cm^{-2}~s^{-1}}$. This shows that the abundance
  ratios in the accreted matter can be calculated once the diffusion
fluxes for the elements are known. The diffusion fluxes are thus the
final parameters in the abundance analysis, replacing the usual
photospheric values. As a consequence, in the next step we have
calculated synthetic spectra with stratified abundances, and adjusted
these fluxes until the spectra are fit satisfactorily. When using the
well known terms of gravitation, electric field, and concentration
gradient in the diffusion equation, the abundances near optical depth
2/3 do not change very much (typically $<$0.1~dex) and the influence
on the derived accretion abundances is noticeable, but minor. This
changes drastically, however, if radiative levitation of some elements
becomes important.

\section{Diffusion and radiative levitation\label{s-raddiff}}
Radiative levitation, the selective acceleration of individual ions by
transfer of momentum from photons, has been studied extensively in the
past \citep[e.g.][and references therein]{Chayer.Fontaine.ea95,
  Chayer.Vennes.ea95}, but was usually considered important only at
temperatures above 30\,000~K. However, \cite{Chayer.Dupuis10} have
demonstrated that radiative levitation could support Si at a low level
in a DA atmosphere model with $T_\mathrm{eff}$\ = 20\,000~K and $\log
g$\ = 8. Marginal support could also be possible for C.  An
application of this result to four DAZ white dwarfs in the temperature
range 18\,000--25\,000~K \citep{Dupuis.Chayer.ea10} showed that in
three cases the Si abundance could be explained without current
accretion. Very recently, \cite{Chayer14} studied two DA stars in the
Hyades (WD\,0421+162 and WD\,0431+126), which are also in our sample,
and demonstrated that, according to their calculations, the Si in
WD\,0431+126 could be completely supported by radiative levitation
(see however Sect.~\ref{s-hyadeswd}, where our findings differ).

This raises the possibility that the numerous objects with Si
detections in our sample may not currently undergo an accretion
episode, or that at least the derived accretion rate could be
significantly altered by radiative levitation.

We have therefore decided to include this effect in our analysis,
which aims to determine accretion rates from observed photospheric
abundances. We start with the description of diffusion following the
basic equations in \cite{Gonzalez.LeBlanc.ea95} and 
\cite{Vennes.Pelletier.ea88}, with some changes in notation and
simplified for a trace element 2 in background element 1
(i.e. abundance ratio $\gamma\ = 0$)
\begin{eqnarray}
 v_i  =  D_i \ \left[ -\frac{\partial \ln c_{2}}{\partial r}
    + \left( \frac{Z_i}{Z_1} A_1 - A_2 \right) \frac{m g}{kT}\right. \\
    \left. + \left(\frac{Z_i}{Z_1} - 1 \right)\,\frac{\partial 
    \ln p_1}{\partial r} + \frac{A_2\, m\, g_{\mathrm{rad},i}}{kT}
    \right] 
\nonumber
\end{eqnarray}
Here $A_1, A_2$ are the atomic mass numbers of the two elements, $Z_1$
is the average charge of the background, $Z_i$ the charge of ion $i$ of
the trace element, $c_2$ its number fraction, $m$ the atomic mass
unit, $k$ the Boltzmann constant, $T$ temperature, $g$ the
gravitational acceleration, and $p_1$ the partial pressure of the ions
of element 1. $D_i$ and $v_i$ are the diffusion constants and
diffusion velocity for ion $i$ of the trace element.

The radiative force on ion $i$ of the trace element is
\begin{equation}
   A_2\, m\, g_{\mathrm{rad},i} = \frac{1}{c}\,\int_0^\infty \sigma_i
   F_{\lambda}  \ d\lambda = f_{\mathrm{rad},i}
\end{equation}
where $F_{\lambda}$ is the radiative energy flux in the atmosphere,
$\sigma_i$ the line absorption cross section of ion $i$, and $c$
the velocity of light.

There has been some discussion in the literature, starting with
\cite{Gonzalez.LeBlanc.ea95}, about the distribution of the momentum
gained through photon absorption on the different ionization stages
\citep[e.g.][]{Chayer.Fontaine.ea95, Chayer.Vennes.ea95}. The question
is, whether the absorbing ion ionizes to the next higher state before
losing the gained momentum or not. Since the detailed calculations
for each transition are much too complicated to be incorporated into
our present analysis, we follow the simple prescription of
\cite{Chayer.Vennes.ea95}: if the principal quantum number of the final
state of the transition is not higher than that of the ground state
plus one, the momentum is lost in the absorbing ion, otherwise in the
next higher state. This is implemented in the following way: dividing
the radiative force in the two contributions according to the above
criterion (index 1 for the momentum lost by the same ion that has
absorbed it, index 2 for the momentum lost in the next higher
ionization state)
\begin{equation}
f_{\mathrm{rad},i} = f_{1,\mathrm{rad},i} + f_{2,\mathrm{rad},i-1}
\end{equation}
we can with the number density $n$ write the total force on a unit
volume as
\begin{eqnarray}
   n f_\mathrm{rad} &=& \sum_i n_i\, (f_{1,\mathrm{rad},i} + f_{2,\mathrm{rad},i-1}) \\
    &=&  n_1 f_{1,\mathrm{rad},1} + \sum_{i=2} n_i\, (f_{1,\mathrm{rad},i} +
    \frac{n_{i-1}}{n_i} f_{2,\mathrm{rad},i-1}) \nonumber .
\end{eqnarray}
This outlines how the effective radiative acceleration for
each ion should be calculated to preserve the total absorbed
momentum. The final step is the averaging of the velocities, weighting
with the ionization fractions of the trace element,
\begin{equation}
       v = \sum_i \frac{n_i}{n} v_i 
\end{equation}
and of the diffusion flux in $\mathrm{g~cm^{-2}~s^{-1}}$
\begin{equation}
  f_\mathrm{diff} = X \rho\ v 
\end{equation}
with mass density $\rho$ and mass fraction $X$ of the trace element.

We note that all necessary quantities are readily available in the
atmosphere model code during the iteration of the atmospheric
structure; there is no need for any further approximations of the
radiative flux or line profiles. The atmospheric structure and element
distribution are iterated until a completely consistent solution is
obtained for the parameters $T_\mathrm{eff}$, $\log g$, and constant
diffusion flux at all layers.
 
Each spectral line is calculated for at least 21 wavelength points
starting with steps of 1/8 the Doppler width at the line center,
increasing gradually to also cover the broad damping wings. We have
made some test calculations to test the dependence of the result on
the number of spectral lines used. A typical example is the DA model
at $T_\mathrm{eff}$\ = 20\,000~K, $\log g$\ = 8.00 and [Si/H] =
\mbox{-7.0}. With the 66 strongest lines of \ion{Si}{i} to
\ion{Si}{iv} the change in the Si abundance at the same diffusion flux
between calculations with vs. without radiative levitation is 0.23 dex
at $\tau_R = 2/3$, and 0.25 maximum considering all layers in the
model between optical depth $10^{-6}$ and 1000. Using 447 Si lines,
the numbers are 0.25 and 0.29; finally, with a large set of 824 lines
the numbers are 0.23 and 0.32. The major effect is already achieved
with the small line set. If we want to answer the question whether
radiative support without any current accretion is possible at the
lowest abundance levels, we would need very high accuracy of the line
absorption, and thus use as many lines as possible, because the answer
is either yes or no. However, our main emphasis is to determine the
diffusion fluxes, where a change of 0.05 dex is well within the
typical errors. We thus decided to take the medium size set as a
compromise between computing resources and accuracy. The corresponding
calculations for C used 394 \ion{C}{i} to \ion{C}{iv} lines.

\begin{figure}
\centering
\includegraphics[width=0.45\textwidth]{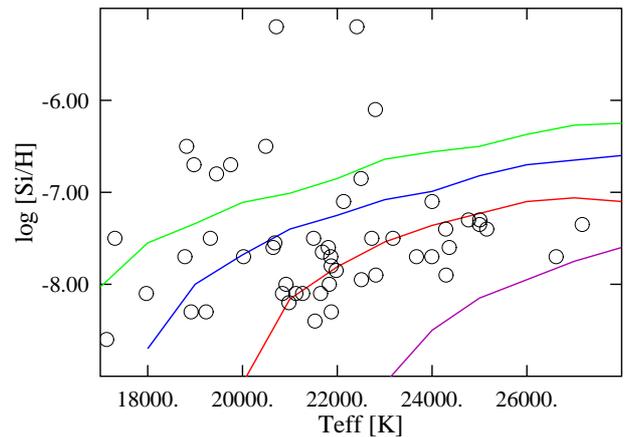}
\caption{Si abundances supported by radiative levitation without
  accretion. Continuous lines are the predictions for surface
  gravities $\log g$\ = 7.50, 7.75, 8.00, 8.25 from top (green, blue,
  red, magenta in the color version). Small circles are the observed
  abundances from Table~\ref{parameters}. The median $\log g$ of the
  stars with observed silicon is 7.94. See text for 
  explanations.} \label{fig5}
\end{figure}

\begin{figure}
\centering \includegraphics[width=0.45\textwidth]{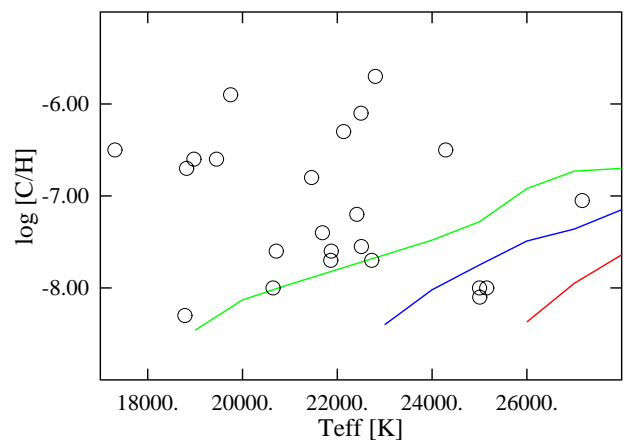}
\caption{Similar to Fig.~\ref{fig5}, but for C. Continuous
  lines are the predictions for surface gravities $\log g$\ = 7.50,
  7.75, 8.00 from top (green, blue, red in the color
  version). The median $\log g$ of the stars with observed carbon
  is 7.90.} \label{fig6}
\end{figure}

To get an overview about the importance of radiative levitation we
have calculated atmospheric models with the element
stratification obtained with zero accretion flow. For these models
synthetic spectra were calculated. In order to compare the results
with the abundances in Table~\ref{parameters} we have then calculated
homogenous models, which produced the same equivalent width as the
stratified models. The abundances of these models are shown in
Fig.~\ref{fig5} and compared to the Si abundances of
Table~\ref{parameters}. Fig.~\ref{fig6} shows the same
comparison for C.

Figs.~\ref{fig5} and \ref{fig6} suggest that for a
significant fraction of objects with photospheric Si radiative levitation
may be important. On the other hand, for most of the C observations
accretion seems to be necessary. The final answer is only possible by
calculating in the next step the levitation for each object
individually, using the appropriate stellar parameters from
Table~\ref{parameters}.

\begin{figure}
\centering \includegraphics[width=0.45\textwidth]{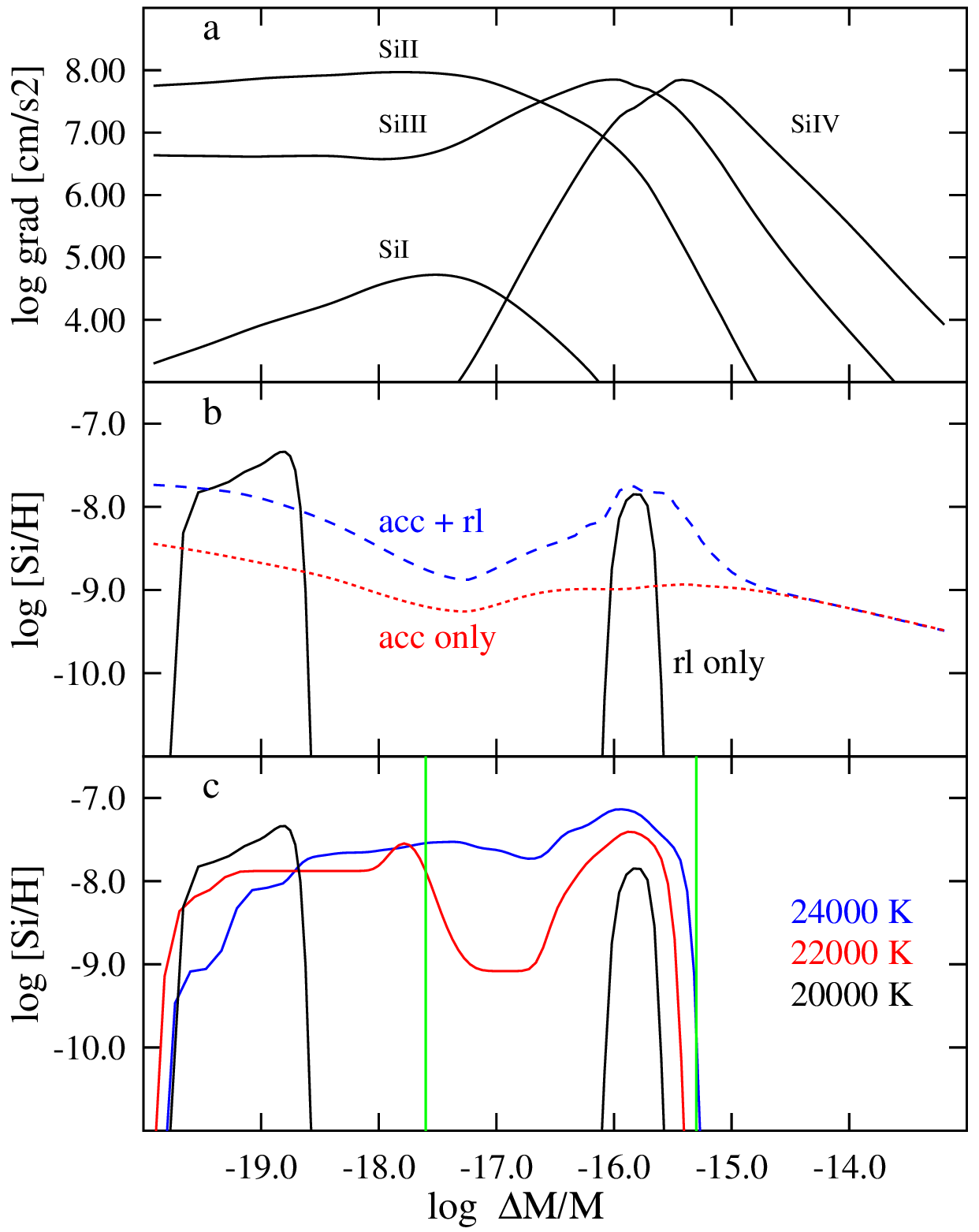}
\caption{Radiative support in a $\log g$\ = 8 DA white dwarf model as
  a function of depth in the atmosphere, expressed as fractional mass
  $\Delta M/M$ from the surface.}
 \label{fig7}
\end{figure}

\subsection{Comparison with previous work}
Radiative levitation in white dwarfs has been calculated in several
  papers of the Montreal group \citep[see e.g.][and references
    therein]{Chayer.Fontaine.ea95, Chayer.Vennes.ea95} and by
  \cite{Dreizler.Wolff99}. These studies were aimed at hot stars. We can,
  however, compare our results with the recent studies of \cite{Chayer14} and
  \citet[][hereafter C10]{Chayer.Dupuis10}. We use the same diffusion
  coefficients from  
  \cite{Paquette.Pelletier.ea86}, but the complete implementation and
  programming is independent. \citet{Chayer.Dupuis10}, give in
their Fig.~3 the abundance distribution of Si in a 20\,000~K, $\log g$\ = 8 DA
model for a range of accretion rates. Our own calculations for the same model
are shown in Fig.~\ref{fig7}. The four panels show:\\
a) Radiative support from the first four ions of silicon at
$T_\mathrm{eff}$\ = 20\,000~K, expressed as acceleration, weighted
with the relative abundances of the ion.\\
b) Si abundance supported by this force. Without accretion two
separate ``clouds'' of Si are formed; outside these clouds the
abundance drops to zero (continuous black line). The dashed (blue)
line shows the Si distribution when a small accretion/diffusion flux
of $3.9\times10^4$ g\,s$^{-1}$ is present, the dotted (red) line shows
the same accretion flux with radiative levitation switched off.\\
c) Si abundance stratification for 24\,000, 22\,000, and 20\,000~K
(from top on the right side). From 24\,000 to the upper limit of our
sample near 27\,000~K, the curves differ very little from the
24\,000~K curve. The main ``visible'' part of the atmosphere between
$\tau_\mathrm{Rosseland}$ 0.01-10 is between the two vertical green
lines.

The maximum abundance supported at the lowest accretion rates in C10
is $\log$ [Si/H] $\approx -7.5$, which is very similar in our own
calculations. However, there are significant differences between the
two approaches
\begin{itemize}
\item C10 solve the time-dependent diffusion equation, whereas we
  determine the stationary state. We are interested in the
  interpretation of spectroscopic observations, which depend on the
  details of the atmosphere; according to Fig.~3 in C10 equilibrium is
  reached in less than one year. Any time-dependent variations
    in the observed DAZ occur very likely on much longer time scales,
    justifying our use of the stationary state.
\item C10 calculate envelope models. The atmospheric conditions
  at optical depth $2/3$ (corresponding to $\log \Delta M/M \approx$-16.4
  in this model) are used as boundary conditions; the Si abundance is
  assumed to be constant above this layer. As the bottom panel of
  Fig.~\ref{fig7} demonstrates, the distribution of Si
  throughout the photosphere is very far from uniform (black
  continuous line). If we aim at a realistic comparison with observed
  spectra, we have to calculate synthetic spectra for this
  distribution, and it is quite clear in this case that the Si lines
  will be much weaker than in a model with constant $\log$ [Si/H] =
  -7.5. We will come back to a specific example below.
\item There is a qualitative difference between a model with zero
  accretion and one with even a very small accretion flux. If
  radiative support fails by a tiny amount to support Si at some layer
  in the atmosphere, the abundance will drop to zero without
  accretion. On the other hand, with accretion, the downward diffusion
  velocity can be very small, because of the near cancellation of the
  force terms in the diffusion equation. This will lead to a strong
  enhancement of the abundances as compared to regions without
  radiative support. This is demonstrated by the (blue) dashed line,
  which shows the Si distribution in the case of a small accretion
  flux ($3.9\times10^4$ g\,s$^{-1}$) and can be compared with the
  dotted (red) line, which shows the same accretion flux, but with
  radiative levitation switched off.
\end{itemize}

\begin{table*}
\caption{Equivalent widths (in m\AA) of \ion{Si}{ii}~1265\,\AA\ and
  \ion{C}{ii}~1335\,\AA\ measured from the observed spectra (obs)
  compared to the predictions of radiative levitation (RL).  
  Logarithms of the C and Si diffusion fluxes (in g\,s$^{-1}$).
  See Sect.~\ref{s-raddiff}, \ref{s-difffluxes} for details.} 
\label{radlev}
\centering
\begin{tabular}{lrrrrllrr}
\hline\hline
\noalign{\smallskip}
 object    & \multicolumn{2}{c}{EW(Si)[m\AA]}  &
 \multicolumn{2}{c}{EW(C)[m\AA]}
           & Notes & R/A & \multicolumn{2}{c}{$\log$ flux [g\,s$^{-1}$]}\\
           &    obs & RL &    obs & RL  &  &  &  Si  &  C \\
\hline
\noalign{\smallskip}
WD0000+171   & 54 & 30 &  0 &  0  &          & R  & <4.20  & <4.02\\
WD0059+257   &291 &  7 & 47 &  0  & 1        & A  &  6.13  & <4.47\\
WD0102+095   & 78 & 63 & 15 & 11  &          & R  & <4.75  & <3.75\\
WD0114-605   & 88 & 80 & 25 & 31  &          & R  & <4.48  & <3.48\\
WD0140-392   & 95 & 41 &  0 &  0  &          & R  & <4.78  & <3.88\\
WD0242-174   & 56 & 16 &  0 &  0  &          & R? & <4.03  & <4.23\\
WD0307+149   & 36 & 29 &  0 &  0  &          & R  & <4.39  & <4.17\\
WD0341+021   & 25 &171 & 82 & 78  &          & R  & <3.39  & <3.94\\
WD0352+018   & 40 & 39 &    &  0  & 5        & R  & <3.82  & <4.08\\
HE0403-4129  & 90 & 36 &    &  0  & 5        & R? & <5.34  & <4.35\\
HE0416-1034  & 33 & 35 &220 & 71  & 6        & A  & <4.70  &  5.28\\
WD0421+162   &144 &  0 &  0 &  0  &          & A  &  5.27  & <4.49\\
WD0431+126   & 55 &  2 &  0 &  0  &          & A  &  4.33  & <4.36\\
HE0452-3444  & 59 & 32 &  0 &  0  &          & R  & <4.61  & <4.20\\
HS0507+0434A & 20 &  5 &  0 &  0  &          & R? & <3.91  & <4.12\\
WD0710+741   &245 & 118&277 & 10  & *,1,2,6  & A  &  5.39  &  6.51\\
WD0843+516   &126 &  0 & 90 &  0  & 1,4      & A  &  7.56  &  5.38\\
WD0854+404   & 41 & 58 &    &  0  & 5,7      & R  & <4.15  & <4.19\\
WD0920+363   & 54 &112 &    & 54  & 5        & R  & <4.35  & <3.68\\
WD0954+697   & 49 & 40 &  0 &  0  &          & R  & <4.08  & <4.23\\
WD1013+256   &159 & 28 &141 &  0  &          & A  &  5.52  &  6.28\\  
WD1015+161   &409 &  0 &  0 &  0  & 1,4      & A  &  6.54  & <4.84\\
WD1017+125   & 98 & 34 &    &  0  & 5        & R? & <4.86  & <4.94\\
WD1034+492   &123 &  0 &  0 &  0  &          & A  &  5.17  & <4.77\\
WD1038+633   & 39 &  3 &    &  0  & 5,7      & A  &  4.69  & <4.59\\
WD1129+155   & 30 &  0 &    &  0  & 5        & A  &  4.55  & <4.76\\
WD1133+293   & 50 & 75 &    &  4  & 5        & R  & <4.23  & <4.38\\
SDSS1228+1040&1090&  0 & 48 &  0  & *,1,4,5  & A  &  7.72  &  5.23\\
HS1243+0132  & 51 & 48 &    &  0  & 5        & R  & <3.19  & <4.00\\
WD1257+048   & 91 & 39 & 38 &  0  &          & A  &  4.80  &  4.71\\
WD1310-305   & 96 & 15 &    &  0  & 5        & A  &  4.70  & <4.51\\
WD1325+279   & 34 & 19 &    &  0  & 5        & R  & <4.25  & <4.36\\
WD1325-089   & 29 &  0 &    &  0  & 5        & A  &  4.20  & <4.75\\
WD1408+323   &121 &  0 & 18 &  0  &          & A  &  4.95  &  4.42\\
WD1451+006   & 27 & 54 &    & 32  & 5        & R  & <4.01  & <3.02\\
WD1548+149   & 69 & 40 & 61 &  0  &          & A  &  4.66  &  5.09\\
WD1647+375   &339 & 59 &218 &  0  & 1        & A  &  6.44  &  7.06\\
WD1713+332   & 37 &145 & 58 & 60  &          & R  & <3.45  & <4.26\\
WD1755+194   & 71 & 72 & 19 & 20  &          & R  & <4.77  & <3.61\\
WD1914-598   & 19 & 12 &    &  0  & 5        & R  & <3.97  & <4.31\\
WD1929+012   &1559& 41 &130 &  0  & *,1,4    & A  &  8.03  &  5.85\\
WD1943+163   & 271&  7 &139 &  0  & 1        & A  &  5.84  &  6.10\\
WD1953-715   & 314&  0 &141 &  0  & 1        & A  &  6.06  &  6.16\\
WD2046-220   &  76& 56 &    &  0  & 5        & R  & <4.46  & <4.52\\
WD2058+181   & 156&  0 &135 &  0  & 1        & A  &  5.37  &  6.28\\
WD2134+218   &  67&  0 &    &  0  & 5        & A  &  4.89  & <4.70\\
HS2229+2335  & 350&  0 &125 &  0  &          & A  &  6.35  &  6.09\\
HE2231-2647  &  36& 48 &    &  0  & 5        & R  & <3.88  & <3.98\\
HS2244+2103  &  32& 45 &    &  0  & 5        & R  & <4.35  & <3.95\\
WD2256+249   & 145& 74 &168 &  0  & *,1,2    & A  &  5.49  &  6.03\\
WD2257+162   &  66&141 &110 & 74  & *,2      & R  & <3.48  & <5.58\\ 
WD2306+124   & 120&  0 & 20 &  0  &          & A  &  4.98  &  4.70\\
WD2322-181   &  84& 23 & 47 &  0  &          & A  &  4.67  &  4.86\\
\hline
\end{tabular}
\tablefoot{ *: ancillary target, not used in sample statistics; 1:
  additional elements detected; 2: post-common envelope (close) binary;
  3: spatially resolved (wide) binary; 4: exhibits excess
  infrared emission; 5: \ion{C}{ii}~1335 line contaminated by
  interstellar line; if field is empty, the spectrum is consistent
  with a photospheric EW~=~0; 6: \ion{C}{iii} lines near
  1175\,\AA\ used instead of 1335\,\AA; 7:
  \ion{Si}{ii}~1298\,\AA\ line used instead of 1265\,\AA}
\end{table*}

\section{Determination of diffusion fluxes for the sample with
  photospheric spectral lines \label{s-difffluxes}} 

Since radiative levitation depends strongly on the stellar parameters
-- especially the surface gravity -- and the abundances, we need to
study each object individually. For this purpose we calculated two
sets of models for each set of atmospheric parameters; the first one
without accretion, assuming equilibrium between radiative levitation
and gravitational settling. For the resulting atmospheric structure a
synthetic spectrum was calculated and from this the equivalent widths
of the strong lines \ion{Si}{ii}~1265\,\AA\ and
\ion{C}{ii}~1335\,\AA~--~and in a few cases additional lines~--~were
calculated and compared with observed values. This includes obviously
objects where no radiative support is possible, resulting in zero
equivalent widths. The second set included accretion. The parameter we
use in this calculation is the accretion flux (which in stationary
state is also the diffusion flux). The flux for Si and C was varied,
until a satisfactory fit was achieved; in the case of significant
radiative support we determined an upper limit to the
flux. Technically we used the original abundances from the fit with
homogeneous models as a starting point. Assuming that they are
approximately representative for the layer $\tau = 2/3$, they lead,
with the diffusion velocities calculated for this layer to a starting
value for the flux.

Results of these calculations and the comparison with the observations
are presented in Table~\ref{radlev}. The column labeled R/A is our
assessment, whether accretion is necessary to explain the observed
line strengths or not. Unfortunately there is no simple and completely
objective method for this decision. The strong dependence of radiative
levitation on stellar parameters, and the unavoidable uncertainties of
these parameters, as well as uncertainties of the equivalent widths
for very weak lines and noisy spectra, lead to an intermediate range
without unambiguous decisions. We have aimed to be conservative and
rather err on the assignment of R than A, that is, we want to be
confident that the latter objects are really accreting. If the predicted
EWs are within a factor of two of the observed ones, we consider pure
radiative support (without accretion) as possible, whereas with a
factor of four or larger we conclude ongoing accretion. In between, the
decision was made using the quality of the spectra, and if no decision
seemed possible, the assignment was R?.

With this classification, we conclude that of the 48 objects showing
photospheric Si, 23 must be currently  accreting. Likewise, of the 19
objects with C, 14 must be accreting.  There is one single object in
Table~\ref{radlev} (HE\,0416-1034), where the Si abundance could be
explained by radiative levitation alone, but the C abundance
apparently demands accretion. Other results of Table~\ref{radlev} are
further discussed in Section~\ref{discussion}.

There has been a recent claim by \cite{Deal.Deheuvels.ea13} that the
  diffusion fluxes would change by orders of magnitude, if the thermohaline
  instability would be taken into account. In \cite{Xu.Jura.ea14} we discuss
  some arguments, why we do not think that these calculation apply to the
  diffusion in white dwarfs. In that paper, we were presenting results for
  white dwarfs with an outer convection zone. It is conceivable that at the
  bottom of the convection zone a boundary layer develops with a discontinuity
  in molecular weight, if the accretion of a large amount of external matter
  (of asteroid size) is very fast, compared to diffusion timescales. In the
  current sample the stars do not have any convection zone and we believe that
  the \cite{Deal.Deheuvels.ea13} calculations are irrelevant in this case. A
  more detailed discussion will be given in a future paper (Koester, in
  preparation).

\subsection{Effect of parameter errors on diffusion calculations}
We use the same object as above (WD\,1943+163) to estimate the change
in the diffusion fluxes within the error estimates for
$T_\mathrm{eff}$ and $\log g$. With unchanged abundances, the
diffusion flux changes for all elements by approximately 0.05 dex, in
the same direction that the abundances have to change in the spectral
analysis. So the combined uncertainty of the atmospheric parameters on
the final diffusion fluxes can be estimated as 0.09 dex. This should
be added quadratically to the (statistical) abundance uncertainties to
get errors for the individual fluxes. However, since the changes go
into the same direction for all elements, these errors need
not be added when calculating element ratios.

\section{Results and Discussion \label{discussion}}
\begin{figure*}
\centering \includegraphics[width=0.9\textwidth]{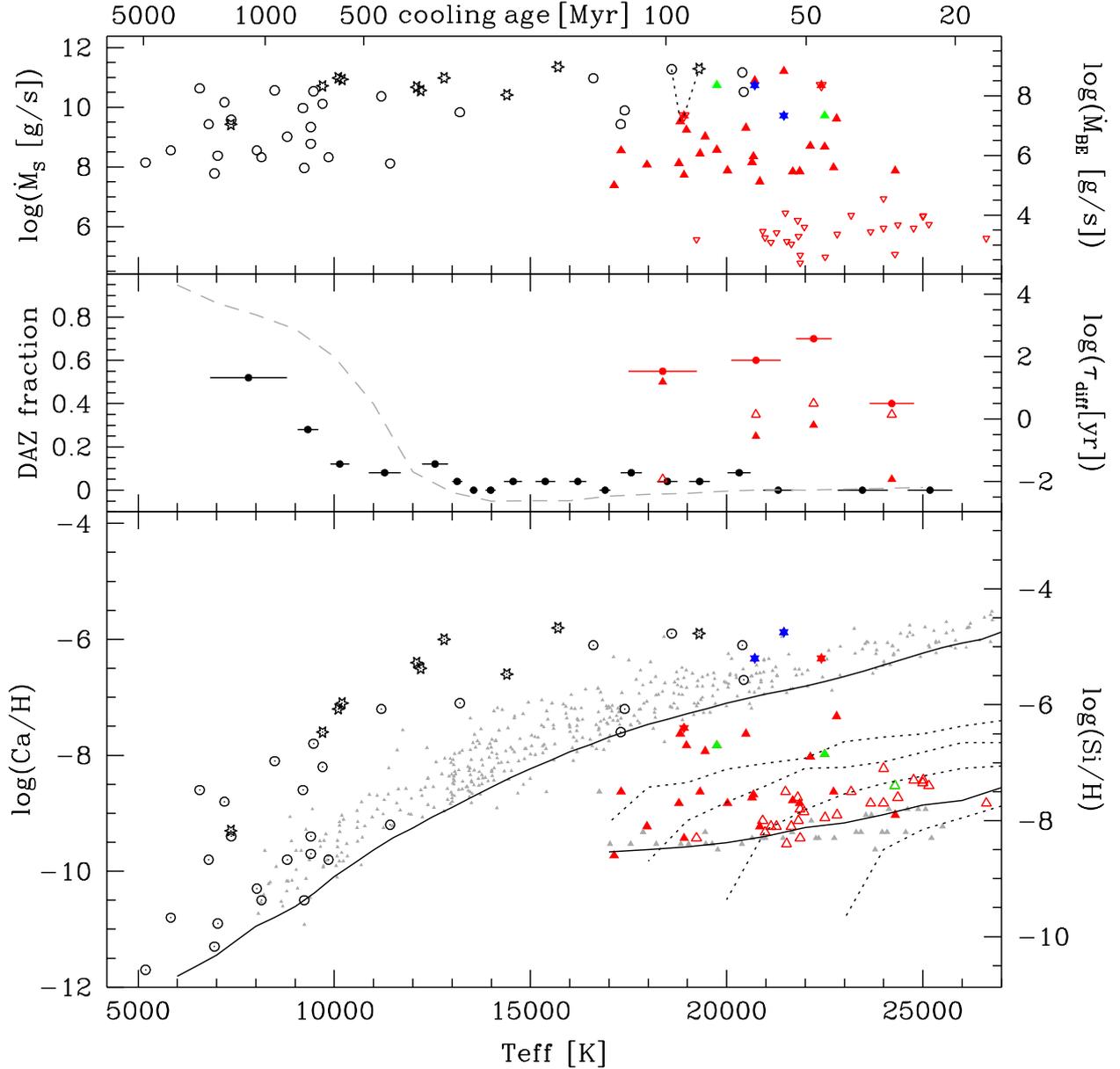}
\caption{Metal pollution in DA white dwarfs as a function of
  $T_\mathrm{eff}$\ (bottom axis) and cooling age (top axis, for
  $\log g$\ =8).  Star-shaped symbols represent white dwarfs where
  their dust discs were detected in the infrared. 
\textbf{Bottom panel:} $\log(\mathrm{Ca/H})$ abundances (open circles and stars)
  and upper limits (gray dots) from \citet{Koester.Wilken06}.  
\textbf{Middle panel:} The fraction of white dwarfs with
  photospheric Ca and Si detections.
\textbf{Top panel:} Inferred total accretion rates, where the Si and Ca
abundances have been scaled by bulk Earth abundances ($\dot M_\mathrm{BE}$)
and by solar abundances ($\dot M_\mathrm{S}$).
See text for further explanations.} 
 \label{fig8}
\end{figure*}

Taken at face value the number of metal-polluted white dwarfs in the
COS sample (48 out of 85, or 56\%, see Table~\ref{abund}) is
surprisingly high.  Previous estimates
(e.g. \citealt{Zuckerman.Koester.ea03, zuckermanetal10-1}) put this
number much lower at 20--30\%. Out of the 48 stars with at least a
detection of photospheric Si, 23 (27\%) must be currently accreting,
and in an additional 25 (29\%), the Si abundance is compatible with
radiative levitation, however, still implying accretion in the recent
past (see Sect.~\ref{s-ismaccretion} below).  This increased fraction
of metal-pollution among our COS sample compared to previous optical
studies is probably due to the relatively high resolution and
signal-to-noise ratio of our observations, and to the choice of the
wavelength region with strong lines of Si, a major component of
planetary debris. Different from previous work on larger samples
\citep{Zuckerman.Koester.ea03, Koester.Rollenhagen.ea05*b,
  Koester.Voss.ea09} our sample was specifically chosen for a search
for metal traces in fairly bright stars in a limited and well defined
temperature range. The detection threshold for Si does not
significantly vary over this temperature range (see the solid
line in the bottom panel of Fig.~\ref{fig8}), and we are confident
that the high fraction of metal-polluted white dwarfs is
representative for the local neighborhood within $\approx$100~pc
around the Sun.

\subsection{Accretion from the interstellar
  medium?\label{s-ismaccretion}} 

Twenty five (29\%) of the stars in our sample have Si abundances that
are consistent with being supported by radiative levitation~--~what is
the nature, and the origin of the polluting material?  Figure~8 in
\cite{Chayer.Vennes.ea95} suggests that the equilibrium abundance of
Si supported by radiative levitation reaches a broad minimum around
$T_\mathrm{eff}$\ $\approx$ 70\,000~K along the cooling sequence for
a 0.6~$M_\sun$ white dwarf (corresponding to a cooling age of
$<1$~Myr). Upon further cooling to 30\,000~K ($\approx$10~Myr), the
abundance necessary for equilibrium rises by a factor of
$\approx$100. At abundances lower than the equilibrium values -- as
those possibly remaining from the earlier evolution -- radiative
levitation will always dominate gravitational settling. Our
interpretation is that any remaining primordial Si would be blown out
of the star by the radiation pressure. As a conclusion, any Si visible
below 30\,000~K must have been accreted.  Before claiming that this
must always be accretion from a remnant planetary system or
circumstellar material in general, we derive some estimates concerning
the possibility of accretion from the interstellar medium.

The total Si mass in radiative levitation equilibrium in the
atmosphere of a 20\,000~K, $\log g=8$ DA is $7.89\times\,10^{10}$~g
(Fig.~\ref{fig7}). The cross section for Eddington accretion
from interstellar matter, which is the minimum we would expect, is
\begin{equation}
   a = \frac{2\pi G M_\mathrm{wd}}{v^2}\,R_\mathrm{wd}  = 
          5.29\times\,10^{22} \mbox{~cm}^2 
\end{equation}
with white dwarf mass and radius $M_\mathrm{wd}$ and $R_\mathrm{wd}$,
gravitational constant $G$ and space velocity $v$. Assuming the white
dwarf crosses just one tiny cloud like our Local Interstellar Cloud
(LIC) with a Si column density of $10^{13}$ cm$^{-2}$ and a velocity
of 30~km\,s$^{-1}$ \citep{Redfield.Linsky04}, it would sweep up
$2.46\times10^{13}$g of Si. That is three hundred times more than
needed to explain the observations, and most of this matter would
diffuse downward, except for the tiny fraction supported by radiative
levitation. However, in this scenario, adopting a typical cloud size
of $\approx$2~pc within the solar neighborhood
\citep{Redfield.Linsky04}, the accretion rate of Si would be
$\dot M$(Si) $<10$~g\,s$^{-1}$, orders of magnitude smaller than the
rates determined for the 23 objects with inferred current accretion. A
more realistic estimate for the accretion of Si bound in interstellar
dust grains is obtained assuming that hydrodynamic (Bondi-Hoyle)
accretion works only within a radius, where the grains sublimate
\citep{Alcock.Illarionov80, Farihi.Barstow.ea10}. A similar calculation
  as above leads to accretion rates $\dot M$(Si) $< 10^3$ g\,s$^{-1}$,
  still much smaller than any of our observed rates.

An additional argument against an ISM origin for the Si detections in
our sample comes from the C/Si ratio that we determined for the debris
material. Accretion from the ISM would suggest a C/Si ratio close to
its solar value, i.e. $\approx$3.6 by mass. Inspection of
Table~\ref{radlev} shows that the majority of the white dwarfs that
are currently accreting have C/Si values, or upper limits thereof,
significantly below solar. Exceptions are the white dwarfs in
post-common envelope binaries, which presumably accrete
$\approx$solar-abundance wind from their companions, plus a handful of
apparently single white dwarfs, the nature of which which will be
discussed elsewhere.

\subsection{Total accretion rates}
Given that typically only one, or at best a few, elements are detected
in the white dwarf photosphere, computing the total accretion must
rely on an assumption regarding the chemical composition of the
accreting material. Historically, solar abundances were adopted within
the context of accretion from the interstellar medium, see
e.g. \citet{Koester.Wilken06} for a discussion of deriving total
accretion rates from the photospheric Ca abundances. Over the past
decade, it has become increasingly clear that planetary debris is the
most likely origin of the metal-pollution detected in a large number
of white dwarfs, and, given that overall the chemical abundance
patterns bear a strong resemblance to rocky solar system material,
using bulk Earth abundances to infer total accretion rates has become
a standard assumption~--~again, most of the times extrapolated from
the photospheric Ca abundances (e.g. \citealt{Farihi.Jura.ea09,
  zuckermanetal10-1, girvenetal12-1}).

Figure~\ref{fig8} summarizes many of our results.  The bottom panel shows
$\log(\mathrm{Ca/H})$ abundances (open circles and stars) and upper limits
(gray dots) from \citet{Koester.Wilken06} (compiled from
\citealt{Zuckerman.Koester.ea03, Berger.Koester.ea05}). Red
symbols illustrate the $\log(\mathrm{Si/H})$ abundances from our HST/COS
survey, where filled red symbols represent the 23 stars that must currently be
accreting (labeled ``A'' in Table~\ref{radlev}), open triangles are the 25
stars where Si is detected close to the equilibrium abundance for radiative
levitation (labeled ``R'' and ``R?'' in Table~\ref{radlev}), and upper limits
to $\log(\mathrm{Si/H})$ for the remaining 37 stars are shown by small gray
triangles. For comparison, the parameters of five additional DA white dwarfs
that we observed with COS \citep{Gansicke.Koester.ea12} are shown in blue (the
debris-accreting SDSS\,1228+1040 and WD\,1929+019) and green (the three PCEBs
WD\,0710+741, WD\,2256+249, WD\,2257+162).  These five stars are not included
in the discussion of the COS DA sample statistics (Sect.~\ref{s-fraction} \&
middle panel of this figure). Infrared excesses from circumstellar dust have
been observed at four stars observed with COS (blue: SDSS\,1228+1040 and
WD\,1929+019, red: WD\,0843+516 and WD\,1015+161). The $\log(\mathrm{Ca/H})$
and $\log(\mathrm{Si/H})$ axes are offset by the bulk Earth
$\log(\mathrm{Si/Ca})=-1.128$ \citep{mcdonough00-1}. The solid lines
correspond to the typical detection limits of Ca achieved by high-resolution
ground-based spectroscopy (EW\,$\approx$15~m\AA\ in the Ca\,K line), and of Si
in our HST/COS observations (EW\,$\approx$30~m\AA\ in Si\,II\,1265\,\AA). The
four dashed lines give the Si equilibrium abundance for $\log g$\ =7.50, 7.75,
8.00, and 8.25 (from top to bottom).

In the middle panel the fraction of white dwarfs with photospheric Ca and Si
detections is shown. Each black dot represents 25 stars with Ca measurements,
with the error bar given by the standard deviation in $T_\mathrm{eff}$. Each
red dot represents 20 stars {\bf with} Si measurements from our COS survey,
which are further sub-divided into white dwarfs that are currently accreting
(filled triangles), and in which photospheric Si could be maintained by
radiative levitation (open triangles). The gray dashed line illustrates the
diffusion time scale for Ca at $\tau=2/3$ in radiative atmospheres, and at the
bottom of the convection zone in convective atmospheres, the transition
between the two regimes occurs between 12\,000~K and 13\,000K.

The top panel of Fig.~\ref{fig8} shows the total accretion rates for the 38
DAZ from \citet{Koester.Wilken06}, for the 25 accreting DAZ from our COS
survey, and for five additional objects that we observed for comparison (two
white dwarfs with infrared excess from their circumstellar debris discs, and
three post-common envelope binaries~--~these objects are not used for the
statistics discussed below). For the stars observed with COS, we scaled the Si
mass fluxes (Table~\ref{radlev}) with the Si mass fraction for bulk-Earth and
solar abundances, 0.161 \citep{mcdonough00-1} and $6.65\times10^{-4}$
\citep{Asplund.Grevesse.ea09}, respectively. Open circles are from Table~3 in
\citet{Koester.Wilken06}, based on the extrapolation from the photospheric Ca
abundances, the filled/colored symbols are extrapolated from the photospheric
Si accretion fluxes determined from our COS spectra.  Open red triangles show
upper limits from our COS survey. Two stars are common to our COS survey and
the \citet{Koester.Wilken06} study (WD1015+161, HS2229+2335), their accretion
rates extrapolated from Ca and Si are joined by dotted lines, and illustrate
the uncertainty in $\dot M$ extrapolated from just one element, given that the
accreting material can show large deviations from the solar or bulk Earth
Si/Ca ratio (see Fig.~7 in \citealt{Gansicke.Koester.ea12}).

While the abundances of planetary debris are broadly speaking
``rock-like'' (I.e. volatile depleted and rich in O, Si, Mg, Fe), we
have shown that there are substantial variations in the metal-to-metal
abundance ratios \citep{Gansicke.Koester.ea12}, which implies that
total accretion rates based on either Ca or Si can easily diverge by
an order of magnitude. WD\,1015+161 and HS\,2229+2335 are both in our
COS sample, and among the DAZ of \citet{Koester.Wilken06}, in both
cases the total accretion rate determined from the photospheric Ca
abundance significantly exceeds that based on our Si measurement.
However, judging from Table~4 and Fig.~7 in
\citet{Gansicke.Koester.ea12}, it is clear that WD\,1015+161 has
overall a much lower Si abundance compared to the other stars. The
case is similar for HS\,2229+2335, where we computed a Ca flux of
$10^{7.15}$~g\,s$^{-1}$ (using $\mathrm{log[Ca/H]} = -5.9$ from
\citealt{Koester.Wilken06}). For white dwarfs with such high
temperatures, ground-based spectroscopy is only sensitive to the
highest Ca abundances, and as such it is maybe not surprising that
both WD\,1015+161 and HS\,2229+2335 have unusually high Ca/Si
ratios. More accurate total accretion rates can only be derived if all
the main elements are detected, i.e. at least O, Mg, Si, and Fe for
rocky material, plus C in the case of volatile-rich material (see
\citealt{Farihi.Gansicke.ea12} for a quantitative comparison).

Taking the accretion rates shown in Fig.~\ref{fig8} at face value,
it is evident that the ranges of rates probed by searches for Ca and Si
at low and high $T_\mathrm{eff}$ are entirely complementary, and
lead to a broadly similar distribution.  Adopting bulk Earth
abundances for the accreted material, the measured accretion rates
range from few $10^5~\mathrm{g~s^{-1}}$, set by the detection
threshold of the observations, to a few $10^8~\mathrm{g~s^{-1}}$. This
upper limit agrees, within an order of magnitude, with the accretion
rate that would be driven purely by Poynting-Robertson drag on dust
particles near the inner edge of the debris disc \citep{rafikov11-1}.

However, sublimation of the dust will unavoidably lead to the
additional presence of gas within the disc, and the gas content may be
enhanced e.g. by the impact of additional asteroids on an existing
debris disc \citep{Jura08}. This gaseous phase has been detected both
from emission lines arising in the outer parts of the disc
\citep{Gansicke.Marsh.ea06, Gansicke.Marsh.ea07,
  Gansicke.Koester.ea08, farihietal12-1, melisetal12-1} and absorption
lines along the line of sight onto the white dwarf
\citep{Debes.Walsh.ea12, Gansicke.Koester.ea12}. The additional
viscosity of this gas is expected to increase the accretion rate over
the value for Poynting-Robertson alone, potentially leading to a
runaway process with peak rates of $10^{10}-10^{11}~\mathrm{g~s^{-1}}$
\citep{rafikov11-2, metzgeretal12-1}. Such high accretion rates are
observationally inferred for a number of DBZ white dwarfs
\citep{Farihi.Gansicke.ea12, girvenetal12-1}. However, because of the
long diffusion time scales in these stars, they are most likely not in
accretion-diffusion equilibrium, and the accretion rates derived for
them should be interpreted as a long-term average value~--~which
implies that the peak accretion rates are probably even higher.  The
absence of any DAZ stars with accretion rates
($\dot{M}>10^{10}$~g\,s$^{-1}$) strongly suggests that such phases
have short life times, and correspondingly small probabilities of
being detected.

A final note concerns the overall distribution of accretion
rates. Ignoring the uncertainties on $\dot M$ for any individual
system, it appears that there is very little dependence of the derived
accretion rates on the cooling age of the stars, which ranges from a
few 10~Myr at the hot end to $\approx$2~Gyr at the cool end. There is
some lack of white dwarfs with low accretion rates in the range
$12\,000~\mathrm{K}\la T_\mathrm{eff} \la 17\,000~\mathrm{K}$, but
this is possibly caused by the decreasing sensitivity of ground-based
spectroscopy for photospheric Ca~H\&K lines. However, at
$T\ga23\,000$~K, there is a sudden drop in stars with high accretion
rates~--~the COS spectroscopy remains extremely sensitive to Si at
these temperatures, so this deficiency is real, and its cause is
discussed below in Sect.~\ref{s-fraction}.

\begin{table}
\caption{Predicted equivalent widths (EW in m\AA) of
  \ion{Si}{ii}~1265\,\AA\ and \ion{C}{ii}~1335\,\AA\ for objects
  where no metal pollution is detected in their COS spectra. In the
  third column Y means the line should be visible, N the
  opposite.}
\label{nosi}
\centering
\begin{tabular}{lrrrrllrr}
\hline\hline
\noalign{\smallskip}
 object    & EW(Si) & EW(C) & Detectable? \\
           &  [m\AA]& [m\AA]&\\  
\hline
\noalign{\smallskip}
 WD0013-241  &   0 & 0 & N\\
 WD0018-339  &  29 & 0 & N\\
 WD0028-474  &   0 & 0 & N\\
 WD0047-524  &   0 & 0 & N\\
 WD0124-257  &  75 &12 & Y\\
 WD0155+069  &  54 & 0 & Y\\
 HS0200+2449 &  13 & 0 & N\\
 HE0308-2305 &   0 & 0 & N\\
 WD0308+188  &   0 & 0 & N\\
 HE0358-5127 &   9 & 0 & N\\
 HE0414-4039 &   7 & 0 & N\\
 HE0418-1021 &   0 & 0 & N\\
 WD0933+025  &  45 & 0 & N\\
 HS0944+1913 &   0 & 0 & N\\
 WD0947+325  &   0 & 0 & N\\
 WD1005+642  &   5 & 0 & N\\
 WD1049-158  &   0 & 0 & N\\
 WD1049+103  &  24 & 0 & N\\
 WD1052+273  &   0 & 0 & N\\
 WD1058-129  &   0 & 0 & N\\
 WD1102+748  &   0 & 0 & N\\
 WD1104+602  &   0 & 0 & N\\
 WD1229-013  & 112 &29 & Y\\
 WD1230-308  &  28 & 0 & N\\
 WD1330+473  &  48 & 0 & Y\\
 WD1353+409  & 128 &60 & Y\\
 WD1459+347  &   0 & 0 & N\\
 WD1524-749  &  91 &30 & Y\\
 WD1531-022  &   0 & 0 & N\\
 WD1547+057  &   0 & 0 & N\\
 WD1633+676  &  34 & 0 & N\\
 WD2021-128  &  21 & 0 & N\\
 WD2032+188  &  59 & 4 & Y\\
 HS2210+2323 &   0 & 0 & N\\
 WD2220+133  &   0 & 0 & N\\
 HE2238-0433 &   0 & 0 & N\\
 WD2359-324  &  70 &22 & Y\\
\hline
\end{tabular}
\end{table}

\begin{figure}
\centering \includegraphics[width=0.48\textwidth]{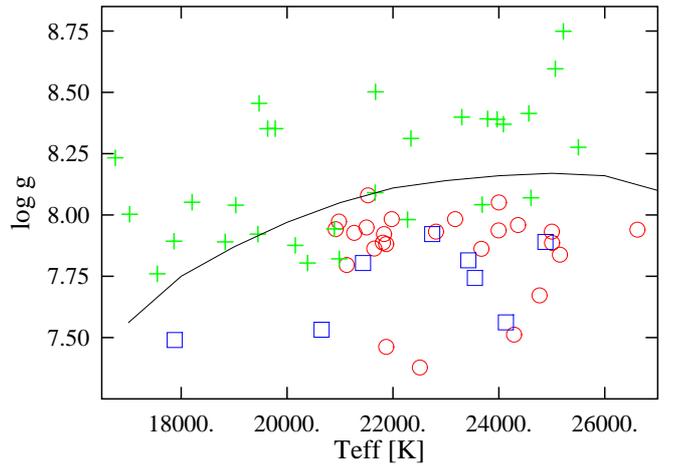}
\caption{Objects where Si is detected at levels consistent with
  radiative levitation (red circles), and objects where Si is neither
  detected, nor expected to be radiatively supported for their
  $T_\mathrm{eff}$ and $\log g$ (green crosses). The blue squares
  indicate the parameters of eight stars where radiative levitation is
  strong enough to produce visible lines, yet, no metals are detected
  in their COS spectra.  The black continuous line indicates the
  location for an EW of 30m\AA\ of the \ion{Si}{ii}~1265 line, which
  is our detection threshold.} \label{fig9}
\end{figure}

\subsection{Objects without photospheric Si}
Radiative levitation predicts detectable amounts of Si over most of the
temperature range of our sample. Why do 37 stars show no trace of Si or C? In
Table~\ref{nosi} we show the predicted equivalent widths of the
\ion{Si}{ii}~1265\,\AA\ and \ion{C}{ii}~1335\,\AA\ lines. The fourth column
(Detectable? N/Y) is our estimate if these predicted lines should be visible
or not, given the individual quality of the COS spectra. We would expect to
detect metals only in 8 of the objects. Two might indeed show some Si, but the
spectrum is strongly perturbed near the 1265\,\AA\ line (WD\,1229-013,
WD\,1230-308).  For the six remaining stars, there is no obvious reason
  why no metals are detected, and our models indicate these systems lack a
  source of external pollution.  Even in the case where 100\% of stars form
  planetary systems, this alone is insufficient to deliver Si to the surfaces
  of all their white dwarf remnants within the appropriate timescale.
  Successful models that deliver debris to the surfaces of white dwarfs
  indicate that a combined planet-planetesimal belt is necessary
  \citep{bonsoretal11-1, Debes.Walsh.ea12} and it seems plausible that
    this architecture may be common but not universal.

Closer inspection of the atmospheric parameters reveals a significant
difference between the stars where no metals are detected, and those
where metals are observed at abundances compatible with radiative
levitation (Table~\ref{radlev}), which is illustrated in
Fig.~\ref{fig9}: the latter objects are concentrated in the lower
right corner, at high temperatures and low surface gravities. Given
the mechanism of radiative levitation this is exactly what is
expected, in objects with lower temperatures and/or higher gravity, Si
sinks out of the photosphere. The continuous line indicating the
combination of $T_\mathrm{eff}$ and $\log g$ where radiative
levitation results in 30~m\AA\ equivalent widths, which is our
typical detection threshold, neatly fits to the division of the two
sets of stars. This is a strong indication that our calculations are
overall correct.

\subsection{The fraction of white dwarfs with remnants of planetary
  systems\label{s-fraction}} 

While the total fraction of Si-polluted white dwarfs in our sample is
relatively independent of $T_\mathrm{eff}$, hovering at $\approx$50\% (red
points in the middle panel of Fig.~\ref{fig8}), the distribution of
white dwarfs that must be currently accreting and those where the
metal pollution can be explained by the equilibrium abundance obtained
from the radiative levitation calculations show a very distinct
pattern. At temperatures below $\approx$20\,000~K, radiative levitation
becomes very ineffective, and correspondingly the vast majority of
stars with Si detections must be currently accreting~--~in other
words, the fraction of accreting white dwarfs is equal to the fraction
of stars with Si detections, $\approx$50\%. Towards higher temperatures,
this pattern reverses, and the majority of Si detections are
consistent with radiative levitation alone. In addition, as already
noted in the previous section, there is a striking lack of stars with
high accretion rates at $T_\mathrm{eff}\ga23\,000$~K.

As argued in Sect.~\ref{s-ismaccretion}, planetary debris is the only
plausible source of metals for the 25 stars that are currently
accreting. Looking at Fig.~\ref{fig8}, these stars have cooling ages
of $\approx$40--100~Myr, in contrast, at younger cooling ages we find
only stars with relatively low photospheric Si abundances which can be
explained by radiative levitation. Yet, as shown in
Sect.~\ref{s-ismaccretion}, these stars also must have accreted at
some point in the past.  A priori, and dynamically, there is no reason
why ongoing accretion of planetary debris should become suddenly more
frequent at white dwarfs with cooling ages older than 40~Myr.

One plausible explanation for this dichotomy relies on the fact that
observing a white dwarf in a phase of ongoing accretion
  requires the availability of a reservoir of material. Such a
reservoir has been found in the form of dusty and gaseous discs around
30 white dwarfs \citep[e.g.][]{Zuckerman.Becklin87,
  Gansicke.Marsh.ea06, Farihi.Jura.ea09, brinkworthetal-12}. While
strongly constrained only for about 1/3 of known dusty white dwarfs
that were observed with SPITZER out to 24~$\mu$, the outer radii of
these discs are compatible with the tidal disruption radius (Roche
limit) of a typical white dwarf, $\approx$1~$R_\sun$. Their inner
radii are consistent with the distance from the white dwarf at which
the radiation field is sufficiently strong to sublimate the
dust. Hence, the location of the inner disc radii depends on the
composition of the dust and its typical grain sizes (which determines
the sublimation temperature), and to a greater extent on the white
dwarf luminosity. While the dust parameters are somewhat uncertain, it
is straightforward to see that the inner edge of the disc moves out
with increasing $T_\mathrm{eff}$.  Von Hippel et
al. (\citeyear{von-Hippel.Kuchner.ea07}) and \cite{Farihi11} showed
that, adopting a range of assumptions on grain size and sublimation
temperature, the inner (sublimation) radius of a dusty debris disc
will become equal to the Roche limit somewhere in the range
15\,000--22\,000~K. More recently, \cite{rafikov+garmilla12-1} showed
that the sublimation temperatures of Si-rich minerals in a H-deficient
environment are a few 100~K higher than the typically adopted values
(valid in H-rich conditions, such as protoplanetary discs), suggesting
that the upper end of that temperature range is more realistic.

The implication of all the above is that while planetesimal disruption
events can and probably do occur as frequent at white dwarfs hotter
than $\approx$23\,000~K, and dynamically this is actually favored at
earlier cooling ages \citep{Debes.Sigurdsson02, Bonsor.Wyatt11,
  Veras.Mustill.ea13, Frewen.Hansen14}, the debris will rapidly
sublimate, forming a purely gaseous disc. Given that the viscosity of
gas is much higher than that of dust, angular momentum transfer in a
gaseous disc is substantially more efficient, dramatically reducing
the life time of a gaseous disc compared to the dust discs that form
around cooler white dwarf. Once the debris of the tidal disruption
event is delivered onto the white dwarf, the small amount of Si in the
photosphere, sustained by radiative levitation, remains as
unmistakable evidence that the star has undergone an episode of
accretion.  Our hypothesis is corroborated by the fact that no white
dwarfs with close-in dusty discs have been found at
$T_\mathrm{eff}\ga23000$~K \citep{Farihi.Jura.ea09, steeleetal11-1,
  girvenetal11-1, xuetal13-2, girvenetal12-1,
  barberetal12-1}.\footnote{The hottest white dwarf hosting a close-in
  dusty disc is WD\,0843+516, with $T_\mathrm{eff}=22412\pm304$~K
  (Table~1, see also \citealt{Gansicke.Koester.ea12} and
  \citealt{xu+jura12-1}). For completeness we note that infrared
  excesses of $T \approx$100~K material have been detected around
  several hot and young white dwarfs that are central stars of
  planetary nebulae \citep{suetal07-1, chuetal09-1, chuetal11-1,
    bilikovaetal12-1}. However this material is inferred to lie at
  much larger separations from the star, at many tens of AU and its
  nature is still uncertain.}

In summary, at least 27\% of the white dwarfs in our sample are currently
accreting, and another 29\% have photospheric metals sustained by radiative
levitation. In the vast majority of the stars where both Si and C are
detected, the C/Si ratio suggests planetary material as the origin of the
material. For the accreting white dwarfs, this is the only plausible origin as
interstellar accretion cannot provide the observed accretion rates. Also
taking the distribution of accreting white dwarfs into account, along with
those where radiative levitation is sufficient to explain the detected metals,
strongly suggests that the majority of all the metal-polluted stars in our COS
sample are accreting, or have accreted planetary debris.

\subsection{Dependence on the mass of the white dwarfs and their
  progenitors} 

The vast majority of exo-planet searches have focused on FGK type host
stars, corresponding to a very narrow range in stellar masses,
$\approx$0.5--1.5~$M_\sun$\footnote{Search for radial velocity variations
  and planetary transits becomes increasingly difficult with
  increasing mass: A-type stars have very few sharp features in their
  spectra that can be used for cross-correlation, and the depth of
  transits scales as $R^2_\mathrm{p}/R^2_*$, with $R_\mathrm{p}$ and
  $R_*$ the planet and host star radii. The former issue can be
  circumvented by observing ``retired A-stars'', i.e subgiants with
  estimated $M>1.5$~$M_\sun$. While there have been a number of planet
  detections, there is an intense ongoing debate regarding the true
  masses of subgiant planet hosts \citep{bowleretal10-1,
    johnsonetal10-1, johnsonetal13-1, lloyd13-1, lloyd11-1}.
}. Consequently, our current knowledge of the frequency, and
architecture of planet hosts with $M>1.5$~$M_\sun$\ is limited.

White dwarfs, on the other hand, are the burnt-out cores of stars with
initial masses in the range $\approx$0.8--8~$M_\sun$, and the progenitors
of the majority of the present-day galactic white dwarf population had
$\approx$2~$M_\sun$. The white dwarf mass relates to the mass of
its progenitor via the initial-mass to final-mass relation
\citep{Weidemann84, Weidemann00, Catalan.Isern.ea08, kaliraietal08-1,
  Williams.Bolte.ea09, Dobbie.Napiwotzki.ea09}. The distribution of
the progenitor masses of the 85 DA white dwarfs in our sample is shown
in Fig.~\ref{fig1}, illustrating that the median progenitor
of our sample had indeed $\approx$2~$M_\sun$.

The detection of debris-pollution hence probes the existence of
planetary bodies around main sequence stars spanning a wide range of
masses. More than half of the 23 stars in our sample that are
currently accreting have estimated progenitor masses in excess of
2~$M_\sun$, including the two Hyades white dwarfs discussed by Farihi
et al. (\citeyear{Farihi.Gansicke.ea13}, see also
Sect.~\ref{s-hyadeswd}) and WD\,1015+161 which also exhibits infrared
excess from circumstellar dust \citep{Jura.Farihi.ea07}. This
unmistakably demonstrates that the formation and existence of rocky
planetary material around A-stars stars is common, and can survive the
post-main sequence evolution of these stars. This is further
strengthening the results of \citet{bonsoretal14-1} who detected cool
dust with HERSCHEL around 11~\% of sub-giants with estimated
masses in the range 1.5--1.8~$M_\sun$ (see previous footnote).

Intriguing is an apparent lack of debris pollution for the highest
mass white dwarfs, $M_\mathrm{wd}>0.8$~$M_\sun$, corresponding to
progenitors with $M_{ms}\ga3.8$~$M_\sun$~--~we detected Si only in one
out of 14 stars in this mass range, WD\,1038+633 with an estimated
progenitor mass of $\approx$4.3~$M_\sun$.  Given that our COS sample
is still relatively small, we investigated the possible effect of
small-number statistics causing the observed distribution. A
Monte-Carlo simulation, randomizing the white dwarf masses in our
sample, suggests that the lack of debris-polluted white dwarfs with
$M_\mathrm{wd}>0.8$~$M_\sun$\ being a chance result is $\la5$\%; good
enough to justify exploring possible causes for this deficiency.

The white dwarf mass distribution in Fig.~\ref{fig1}
resembles closely that of larger well-studied samples
\citep{Finley.Koester.ea97, Liebert.Bergeron.ea05,
  giammicheleetal12-1}, displaying besides the dominant population
around a mean mass of 0.6~$M_\sun$\ two additional peaks near
$\approx$0.4~$M_\sun$\ and $\approx$0.8~$M_\sun$. We come back to the
low-mass peak below. Regarding the tail of high-mass white dwarfs,
general consensus is that their number is in excess to expectations
from the galactic star formation history, and that a substantial
number of them are the product of double white dwarf mergers. The
complex, and in part violent evolution of such systems makes it
unlikely that planetary bodies present around one, or both of the
progenitors would survive, explaining the lack of debris-pollution
among those stars. 

Given that the main-sequence life time rapidly increases with
decreasing stellar mass, the Galaxy is not old enough to have formed
low mass white dwarfs ($M_\mathrm{wd}\le0.45$~$M_\sun$) from
single-star evolution. In other words, all low-mass white dwarfs must
have undergone binary interactions that truncated their core-growth
prior to the onset of He-burning, and consequently they have
He-cores. A frequent pathway to He-core white dwarfs is the common
envelope interaction with a main-sequence star, often of low mass
\citep{rebassa-mansergasetal11-1}. In fact, two of the post-common
envelope binaries that we included in our survey as comparison objects
(WD\,0710+741, WD\,2257+162) have He-core white dwarfs. Alternatively,
two sufficiently massive main-sequence stars can evolve into close
double-degenerate binaries, containing one, or two He-core white
dwarfs. There is, however, also a small number of apparently single
low-mass white dwarfs, and \citet{Nelemans.Tauris98} suggested that
massive planets may be sufficient to result in the ejection of the
envelope of their host star.

Our COS sample includes five white dwarfs with masses
$<0.45$~$M_\sun$, which we consider a conservative limit for
containing He-cores. WD\,0341+021 and WD\,1713+332 show photospheric
Si compatible with radiative levitation, whereas no metals are
detected in WD\,1229-013, WD\,1353+409, and WD\,2032+188. All five
systems have been investigated for radial velocities, WD\,1713+332 and
WD\,2032+188 are confirmed double-degenerate binaries with periods of
1.12~d and 5.08~d \citep{Marsh.Dhillon.ea95,
  Nelemans.Napiwotzki.ea05}, WD\,0341+021 is a suspected
double-degenerate \citep{Maxted.Marsh.ea00}, and WD\,1229-013 and
WD\,1353+409 show no radial velocity variations \citep{Maxted.Marsh98,
  Maxted.Marsh.ea00}. Given our arguments in
Sect.~\ref{s-ismaccretion}, it hence appears that stellar wind
accretion can occur onto white dwarfs in close binaries. One other
known example of a very likely He-core that is metal-polluted and has
a dusty debris disc is SDSSJ\,155720.77+091624.7, suggesting that
complex planetary systems in close binary systems can form
\citep{girvenetal11-1, steeleetal11-1, farihietal12-1}.

\subsection{The two Hyades white dwarfs\label{s-hyadeswd}}
\citet{Farihi.Gansicke.ea13} interpreted the detection of photospheric
Si in two Hyades white dwarfs, WD\,0421+162 and WD\,0431+126, as
evidence for the presence of rocky planetesimals. The calculations of
the diffusion fluxes in that study did not account for the effect of
radiative levitation, which attracted some criticism from
\cite{Chayer14}. He finds that WD\,0421+162 needs accretion to explain
the observed Si, but that the WD\,0431+126 can be explained by
radiative levitation alone. While we agree on the first object, we
disagree on the second. In our calculations accretion has to also be
invoked for this object. In our model for zero accretion, we find
indeed radiative support between optical depths $5.1\times10^{-6}$ to
$3.2\times10^{-4}$ with a maximum Si abundance of $\log$ [Si/H] =
$-7.47$, and between 1.9 to 8.4 with a maximum of $-7.72$ (note that
the parameters are similar to those of the model in
Fig.~\ref{fig7}). These numbers are very similar to Chayer's
values in his Fig.~2. However, between these two ``Si clouds'' the
abundance falls below $-12.00$, and the equivalent width for
\ion{Si}{ii}~1265 calculated with this model is only 2~m\AA, as
compared to the observed 55~m\AA. The reasons for the difference are
discussed above; the main point here is that C10 do not calculate the
abundance distribution within the visible photosphere.  In any case,
WD0431+126 is close to the limiting line of Fig.~\ref{fig5}
and small differences in the models can shift the result either way.

In summary, including the effects of radiative levitation somewhat
reduces the diffusion flux in WD\,0431+126 with respect to the value
in \citet{Farihi.Gansicke.ea13}, however the discussion and conclusion
of that paper remain unaffected.

\section{Conclusion}
We have carried out an ultraviolet high-resolution survey of 85 DA
warm $17\,000~\mathrm{K}< T_\mathrm{eff} \la 27\,000$~K white dwarfs
with relatively short cooling ages ($20-200$~Myr), and we found
photospheric metal pollution by at least Si in 52\% of our
target stars. The correct interpretation of the data required a full
treatment of radiative levitation, which we have now implemented in
our atmosphere code.

At least 27\% of these stars are currently accreting material that is
consistent with planetary origin, and an additional 29\% have
undergone at least one episode of accretion as we see Si and in some
cases C that is sustained by radiative levitation. At temperatures
$17\,000$~K $ < T_\mathrm{eff} \la 20\,000$~K, the fraction of white
dwarfs accreting planetary debris is $\approx$50\%. While we cannot
exclude that accretion from the ISM is responsible for the metals in a
few of the stars that are not currently accreting, their distribution
in $T_\mathrm{eff}$ strongly suggest that planetary debris is also
the principle, if not sole, cause of their photospheric
pollution. Hence, the fraction of white dwarfs accreting planetary
debris is plausible close to 50\%, similar to the fraction of stars
hosting super-Earths \citep{Gaidos13}.

At stellar temperatures $\ga23\,000$~K the fraction of currently
accreting white dwarfs rapidly drops, which is consistent with the
fact that these stars are too hot to allow the formation of
long-lived dusty discs that are necessary to provide a continuous flow
of metals. Tidal disruption of planetary bodies in these systems will
lead to short-lived accretion events from purely gaseous discs,
reducing the probability of catching these stars during this
phase~--~yet, radiative levitation provides the memory to past
accretion. One prediction of this hypothesis is that a small number of
warm DA stars should be found to be accreting at much higher rates than the
cooler stars with dusty discs. We note that photospheric and
circumstellar metals are detected in a number of white dwarfs that are
much hotter, and much younger than our sample. The origin of the
metals has been intensively debated in the past
\citep{Shipman.Provencal.ea95, Bannister.Barstow.ea03, Barstow.Good.ea03,
  Dickinson.Barstow.ea12,Dickinson.Barstow.ea12*b}, but it is very
likely that it is, at least in part, also related to planetary debris
\citep{Barstow.Barstow.ea14}.  One particularly promising candidate of a
hot white dwarf accreting planetary debris at a high rate is the
extremely metal-polluted GD394 ($\approx$40\,000~K,
\citealt{barstowetal96-1, dupuisetal00-1}).

We find that about half of the white dwarfs that are currently
accreting planetary debris have progenitor masses of 2--3~$M_\sun$
corresponding to late B- and A-type stars, which shows that the
formation of rocky planetary material is common around main-sequence
stars in this mass range. At the highest white dwarf masses,
$M_\mathrm{wd}>0.8$~$M_\sun$, metal pollution is extremely uncommon,
consistent with the hypothesis that most of these massive white dwarfs
are the product of double-degenerate mergers, instead of single-star
evolution.

Finally, combining published ground-based searches for metal-pollution
in white dwarfs with the results from our COS survey, we find that
neither the fraction of accreting white dwarfs, nor the rates at which
they accrete, show a noticeable correlation over an extremely wide
range of cooling ages, $\approx$20~Myr--2~Gyr. This broadly confirms
the evolutionary simulations of \citet{Veras.Mustill.ea13} who found that
instabilities in the remnants of two-planet systems are likely to
occur over roughly the same period in time.

\begin{acknowledgements}
Based on observations made with the NASA/ESA Hubble
Space Telescope, obtained at the Space Telescope Science Institute,
which is operated by the Association of Universities for Research in
Astronomy, Inc., under NASA contract NAS 5-26555. These observations
are associated with program \#12169 and \#12474. 
The research leading to these results has received funding from the
European Research Council under the European Union's Seventh Framework
Programme (FP/2007-2013) / ERC Grant Agreement n. 320964 (WDTracer).
BTG was supported in part by the UK’s Science and Technology
Facilities Council (ST/I001719/1).  
Balmer lines in the models were
calculated with the modified Stark broadening profiles of
\cite{Tremblay.Bergeron09} kindly made available by the authors.
This research was made possible through the use of the AAVSO Photometric
All-Sky Survey (APASS), funded by the Robert Martin Ayers Sciences Fund. 

\end{acknowledgements}


\begin{thebibliography}{142}
\expandafter\ifx\csname natexlab\endcsname\relax\def\natexlab#1{#1}\fi

\bibitem[{{Aannestad} {et~al.}(1993){Aannestad}, {Kenyon}, {Hammond}, \&
  {Sion}}]{Aannestad.Kenyon.ea93}
{Aannestad}, P.~A., {Kenyon}, S.~J., {Hammond}, G.~L., \& {Sion}, E.~M. 1993,
  \aj, 105, 1033

\bibitem[{{Alcock} \& {Illarionov}(1980)}]{Alcock.Illarionov80}
{Alcock}, C. \& {Illarionov}, A. 1980, \apj, 235, 541

\bibitem[{{Allard} \& {Kielkopf}(1982)}]{Allard.Kielkopf82}
{Allard}, N. \& {Kielkopf}, J. 1982, Reviews of Modern Physics, 54, 1103

\bibitem[{{Allard} {et~al.}(1999){Allard}, {Royer}, {Kielkopf}, \&
  {Feautrier}}]{Allard.Royer.ea99}
{Allard}, N.~F., {Royer}, A., {Kielkopf}, J.~F., \& {Feautrier}, N. 1999, \pra,
  60, 1021

\bibitem[{{Asplund} {et~al.}(2009){Asplund}, {Grevesse}, {Sauval}, \&
  {Scott}}]{Asplund.Grevesse.ea09}
{Asplund}, M., {Grevesse}, N., {Sauval}, A.~J., \& {Scott}, P. 2009, \araa, 47,
  481

\bibitem[{{Bannister} {et~al.}(2003){Bannister}, {Barstow}, {Holberg}, \&
  {Bruhweiler}}]{Bannister.Barstow.ea03}
{Bannister}, N.~P., {Barstow}, M.~A., {Holberg}, J.~B., \& {Bruhweiler}, F.~C.
  2003, \mnras, 341, 477

\bibitem[{{Barber} {et~al.}(2012){Barber}, {Patterson}, {Kilic}, {Leggett},
  {Dufour}, {Bloom}, \& {Starr}}]{barberetal12-1}
{Barber}, S.~D., {Patterson}, A.~J., {Kilic}, M., {et~al.} 2012, \apj, 760, 26

\bibitem[{{Barstow} {et~al.}(2014){Barstow}, {Barstow}, {Casewell}, {Holberg},
  \& {Hubeny}}]{Barstow.Barstow.ea14}
{Barstow}, M.~A., {Barstow}, J.~K., {Casewell}, S.~L., {Holberg}, J.~B., \&
  {Hubeny}, I. 2014, ArXiv e-prints

\bibitem[{{Barstow} {et~al.}(2003){Barstow}, {Good}, {Holberg}, {Hubeny},
  {Bannister}, {Bruhweiler}, {Burleigh}, \& {Napiwotzki}}]{Barstow.Good.ea03}
{Barstow}, M.~A., {Good}, S.~A., {Holberg}, J.~B., {et~al.} 2003, \mnras, 341,
  870

\bibitem[{{Barstow} {et~al.}(1996){Barstow}, {Holberg}, {Hubeny}, {Lanz},
  {Bruhweiler}, \& {Tweedy}}]{barstowetal96-1}
{Barstow}, M.~A., {Holberg}, J.~B., {Hubeny}, I., {et~al.} 1996, \mnras, 279,
  1120

\bibitem[{{Becklin} {et~al.}(2005){Becklin}, {Farihi}, {Jura}, {Song},
  {Weinberger}, \& {Zuckerman}}]{Becklin.Farihi.ea05}
{Becklin}, E.~E., {Farihi}, J., {Jura}, M., {et~al.} 2005, \apjl, 632, L119

\bibitem[{{Berger} {et~al.}(2005){Berger}, {Koester}, {Napiwotzki}, {Reid}, \&
  {Zuckerman}}]{Berger.Koester.ea05}
{Berger}, L., {Koester}, D., {Napiwotzki}, R., {Reid}, I.~N., \& {Zuckerman},
  B. 2005, \aap, 444, 565

\bibitem[{{Bergeron} {et~al.}(2011){Bergeron}, {Wesemael}, {Dufour},
  {Beauchamp}, {Hunter}, {Saffer}, {Gianninas}, {Ruiz}, {Limoges}, {Dufour},
  {Fontaine}, \& {Liebert}}]{bergeronetal11-1}
{Bergeron}, P., {Wesemael}, F., {Dufour}, P., {et~al.} 2011, \apj, 737, 28

\bibitem[{{Bil{\'{\i}}kov{\'a}} {et~al.}(2012){Bil{\'{\i}}kov{\'a}}, {Chu},
  {Gruendl}, {Su}, \& {De Marco}}]{bilikovaetal12-1}
{Bil{\'{\i}}kov{\'a}}, J., {Chu}, Y.-H., {Gruendl}, R.~A., {Su}, K.~Y.~L., \&
  {De Marco}, O. 2012, \apjs, 200, 3

\bibitem[{{Bonsor} {et~al.}(2014){Bonsor}, {Kennedy}, {Wyatt}, {Johnson}, \&
  {Sibthorpe}}]{bonsoretal14-1}
{Bonsor}, A., {Kennedy}, G.~M., {Wyatt}, M.~C., {Johnson}, J.~A., \&
  {Sibthorpe}, B. 2014, \mnras, 437, 3288

\bibitem[{{Bonsor} {et~al.}(2011){Bonsor}, {Mustill}, \&
  {Wyatt}}]{bonsoretal11-1}
{Bonsor}, A., {Mustill}, A.~J., \& {Wyatt}, M.~C. 2011, \mnras, 414, 930

\bibitem[{{Bonsor} \& {Wyatt}(2011)}]{Bonsor.Wyatt11}
{Bonsor}, A. \& {Wyatt}, M.~C. 2011, in American Institute of Physics
  Conference Series, Vol. 1331, American Institute of Physics Conference
  Series, ed. S.~{Schuh}, H.~{Drechsel}, \& U.~{Heber}, 41--48

\bibitem[{{Bowler} {et~al.}(2010){Bowler}, {Johnson}, {Marcy}, {Henry}, {Peek},
  {Fischer}, {Clubb}, {Liu}, {Reffert}, {Schwab}, \& {Lowe}}]{bowleretal10-1}
{Bowler}, B.~P., {Johnson}, J.~A., {Marcy}, G.~W., {et~al.} 2010, \apj, 709,
  396

\bibitem[{{Brinkworth} {et~al.}(2012){Brinkworth}, {G{\"a}nsicke}, {Girven},
  {Hoard}, {Marsh}, {Parsons}, \& {Koester}}]{brinkworthetal-12}
{Brinkworth}, C.~S., {G{\"a}nsicke}, B.~T., {Girven}, J.~M., {et~al.} 2012,
  \apj, 750, 86

\bibitem[{{Cassan} {et~al.}(2012){Cassan}, {Kubas}, {Beaulieu}, {Dominik},
  {Horne}, {Greenhill}, {Wambsganss}, {Menzies}, {Williams}, {J{\o}rgensen},
  {Udalski}, {Bennett}, {Albrow}, {Batista}, {Brillant}, {Caldwell}, {Cole},
  {Coutures}, {Cook}, {Dieters}, {Prester}, {Donatowicz}, {Fouqu{\'e}}, {Hill},
  {Kains}, {Kane}, {Marquette}, {Martin}, {Pollard}, {Sahu}, {Vinter},
  {Warren}, {Watson}, {Zub}, {Sumi}, {Szyma{\'n}ski}, {Kubiak}, {Poleski},
  {Soszynski}, {Ulaczyk}, {Pietrzy{\'n}ski}, \& {Wyrzykowski}}]{cassanetal12-1}
{Cassan}, A., {Kubas}, D., {Beaulieu}, J.-P., {et~al.} 2012, Nature, 481, 167

\bibitem[{{Catal{\'a}n} {et~al.}(2008){Catal{\'a}n}, {Isern},
  {Garc{\'{\i}}a-Berro}, \& {Ribas}}]{Catalan.Isern.ea08}
{Catal{\'a}n}, S., {Isern}, J., {Garc{\'{\i}}a-Berro}, E., \& {Ribas}, I. 2008,
  \mnras, 387, 1693

\bibitem[{{Chayer}(2014)}]{Chayer14}
{Chayer}, P. 2014, \mnras, 437, L95

\bibitem[{{Chayer} \& {Dupuis}(2010)}]{Chayer.Dupuis10}
{Chayer}, P. \& {Dupuis}, J. 2010, in American Institute of Physics Conference
  Series, Vol. 1273, American Institute of Physics Conference Series, ed.
  K.~{Werner} \& T.~{Rauch}, 394--399

\bibitem[{{Chayer} {et~al.}(1995{\natexlab{a}}){Chayer}, {Fontaine}, \&
  {Wesemael}}]{Chayer.Fontaine.ea95}
{Chayer}, P., {Fontaine}, G., \& {Wesemael}, F. 1995{\natexlab{a}}, \apjs, 99,
  189

\bibitem[{{Chayer} {et~al.}(1995{\natexlab{b}}){Chayer}, {Vennes}, {Pradhan},
  {Thejll}, {Beauchamp}, {Fontaine}, \& {Wesemael}}]{Chayer.Vennes.ea95}
{Chayer}, P., {Vennes}, S., {Pradhan}, A.~K., {et~al.} 1995{\natexlab{b}},
  \apj, 454, 429

\bibitem[{{Chu} {et~al.}(2009){Chu}, {Gruendl}, {Guerrero}, {Su}, {Bilikova},
  {Cohen}, {Parker}, {Volk}, {Caulet}, {Chen}, {Hora}, \&
  {Rauch}}]{chuetal09-1}
{Chu}, Y.-H., {Gruendl}, R.~A., {Guerrero}, M.~A., {et~al.} 2009, \aj, 138, 691

\bibitem[{{Chu} {et~al.}(2011){Chu}, {Su}, {Bilikova}, {Gruendl}, {De Marco},
  {Guerrero}, {Updike}, {Volk}, \& {Rauch}}]{chuetal11-1}
{Chu}, Y.-H., {Su}, K.~Y.~L., {Bilikova}, J., {et~al.} 2011, \aj, 142, 75

\bibitem[{{Davidsson}(1999)}]{davidsson99-1}
{Davidsson}, B.~J.~R. 1999, Icarus, 142, 525

\bibitem[{{Deal} {et~al.}(2013){Deal}, {Deheuvels}, {Vauclair}, {Vauclair}, \&
  {Wachlin}}]{Deal.Deheuvels.ea13}
{Deal}, M., {Deheuvels}, S., {Vauclair}, G., {Vauclair}, S., \& {Wachlin},
  F.~C. 2013, \aap, 557, L12

\bibitem[{{Debes} \& {Sigurdsson}(2002)}]{Debes.Sigurdsson02}
{Debes}, J.~H. \& {Sigurdsson}, S. 2002, \apj, 572, 556

\bibitem[{{Debes} {et~al.}(2012){Debes}, {Walsh}, \&
  {Stark}}]{Debes.Walsh.ea12}
{Debes}, J.~H., {Walsh}, K.~J., \& {Stark}, C. 2012, \apj, 747, 148

\bibitem[{{Dickinson} {et~al.}(2012{\natexlab{a}}){Dickinson}, {Barstow}, \&
  {Hubeny}}]{Dickinson.Barstow.ea12}
{Dickinson}, N.~J., {Barstow}, M.~A., \& {Hubeny}, I. 2012{\natexlab{a}},
  \mnras, 421, 3222

\bibitem[{{Dickinson} {et~al.}(2012{\natexlab{b}}){Dickinson}, {Barstow},
  {Welsh}, {Burleigh}, {Farihi}, {Redfield}, \&
  {Unglaub}}]{Dickinson.Barstow.ea12*b}
{Dickinson}, N.~J., {Barstow}, M.~A., {Welsh}, B.~Y., {et~al.}
  2012{\natexlab{b}}, \mnras, 423, 1397

\bibitem[{{Dobbie} {et~al.}(2009){Dobbie}, {Napiwotzki}, {Burleigh},
  {Williams}, {Sharp}, {Barstow}, {Casewell}, \&
  {Hubeny}}]{Dobbie.Napiwotzki.ea09}
{Dobbie}, P.~D., {Napiwotzki}, R., {Burleigh}, M.~R., {et~al.} 2009, \mnras,
  395, 2248

\bibitem[{{Dreizler} \& {Wolff}(1999)}]{Dreizler.Wolff99}
{Dreizler}, S. \& {Wolff}, B. 1999, \aap, 348, 189

\bibitem[{{Dufour} {et~al.}(2010){Dufour}, {Kilic}, {Fontaine}, {Bergeron},
  {Lachapelle}, {Kleinman}, \& {Leggett}}]{Dufour.Kilic.ea10}
{Dufour}, P., {Kilic}, M., {Fontaine}, G., {et~al.} 2010, \apj, 719, 803

\bibitem[{{Dufour} {et~al.}(2012){Dufour}, {Kilic}, {Fontaine}, {Bergeron},
  {Melis}, \& {Bochanski}}]{Dufour.Kilic.ea12}
{Dufour}, P., {Kilic}, M., {Fontaine}, G., {et~al.} 2012, \apj, 749, 6

\bibitem[{{Duncan} \& {Lissauer}(1998)}]{duncan+lissauer98-1}
{Duncan}, M.~J. \& {Lissauer}, J.~J. 1998, Icarus, 134, 303

\bibitem[{{Dupuis} {et~al.}(2010){Dupuis}, {Chayer}, \&
  {H{\'e}nault-Brunet}}]{Dupuis.Chayer.ea10}
{Dupuis}, J., {Chayer}, P., \& {H{\'e}nault-Brunet}, V. 2010, in American
  Institute of Physics Conference Series, Vol. 1273, American Institute of
  Physics Conference Series, ed. K.~{Werner} \& T.~{Rauch}, 412--417

\bibitem[{{Dupuis} {et~al.}(2000){Dupuis}, {Chayer}, {Vennes}, {Christian}, \&
  {Kruk}}]{dupuisetal00-1}
{Dupuis}, J., {Chayer}, P., {Vennes}, S., {Christian}, D.~J., \& {Kruk}, J.~W.
  2000, \apj, 537, 977

\bibitem[{{Dupuis} {et~al.}(1992){Dupuis}, {Fontaine}, {Pelletier}, \&
  {Wesemael}}]{Dupuis.Fontaine.ea92}
{Dupuis}, J., {Fontaine}, G., {Pelletier}, C., \& {Wesemael}, F. 1992, \apjs,
  82, 505

\bibitem[{{Farihi}(2011)}]{Farihi11}
{Farihi}, J. 2011, {White Dwarf Circumstellar Disks: Observations} (Whiley-VCH,
  Berlin), 117--171

\bibitem[{{Farihi} {et~al.}(2010{\natexlab{a}}){Farihi}, {Barstow}, {Redfield},
  {Dufour}, \& {Hambly}}]{Farihi.Barstow.ea10}
{Farihi}, J., {Barstow}, M.~A., {Redfield}, S., {Dufour}, P., \& {Hambly},
  N.~C. 2010{\natexlab{a}}, \mnras, 404, 2123

\bibitem[{{Farihi} {et~al.}(2013{\natexlab{a}}){Farihi}, {G{\"a}nsicke}, \&
  {Koester}}]{farihietal13-2}
{Farihi}, J., {G{\"a}nsicke}, B.~T., \& {Koester}, D. 2013{\natexlab{a}},
  Science, 342, 218

\bibitem[{{Farihi} {et~al.}(2013{\natexlab{b}}){Farihi}, {G{\"a}nsicke}, \&
  {Koester}}]{Farihi.Gansicke.ea13}
{Farihi}, J., {G{\"a}nsicke}, B.~T., \& {Koester}, D. 2013{\natexlab{b}},
  \mnras, 432, 1955

\bibitem[{{Farihi} {et~al.}(2012{\natexlab{a}}){Farihi}, {G{\"a}nsicke},
  {Steele}, {Girven}, {Burleigh}, {Breedt}, \& {Koester}}]{farihietal12-1}
{Farihi}, J., {G{\"a}nsicke}, B.~T., {Steele}, P.~R., {et~al.}
  2012{\natexlab{a}}, \mnras, 421, 1635

\bibitem[{{Farihi} {et~al.}(2012{\natexlab{b}}){Farihi}, {G{\"a}nsicke},
  {Wyatt}, {Girven}, {Pringle}, \& {King}}]{Farihi.Gansicke.ea12}
{Farihi}, J., {G{\"a}nsicke}, B.~T., {Wyatt}, M.~C., {et~al.}
  2012{\natexlab{b}}, \mnras, 424, 464

\bibitem[{{Farihi} {et~al.}(2010{\natexlab{b}}){Farihi}, {Hoard}, \&
  {Wachter}}]{farihietal10-3}
{Farihi}, J., {Hoard}, D.~W., \& {Wachter}, S. 2010{\natexlab{b}}, \apjs, 190,
  275

\bibitem[{{Farihi} {et~al.}(2009){Farihi}, {Jura}, \&
  {Zuckerman}}]{Farihi.Jura.ea09}
{Farihi}, J., {Jura}, M., \& {Zuckerman}, B. 2009, \apj, 694, 805

\bibitem[{{Farihi} {et~al.}(2008){Farihi}, {Zuckerman}, \&
  {Becklin}}]{Farihi.Zuckerman.ea08}
{Farihi}, J., {Zuckerman}, B., \& {Becklin}, E.~E. 2008, \apj, 674, 431

\bibitem[{{Finley} {et~al.}(1997){Finley}, {Koester}, {Kruk}, {Kimble}, \&
  {Allard}}]{Finley.Koester.ea97}
{Finley}, D., {Koester}, D., {Kruk}, J., {Kimble}, R., \& {Allard}, N. 1997, in
  White Dwarfs, Proc. 10th European Workshop on White Dwarfs, ed. J.~{Isern},
  M.~{Hernanz}, \& E.~{Garcia-Berro}, 245

\bibitem[{{Fontaine} {et~al.}(2001){Fontaine}, {Brassard}, \&
  {Bergeron}}]{Fontaine.Brassard.ea01}
{Fontaine}, G., {Brassard}, P., \& {Bergeron}, P. 2001, \pasp, 113, 409

\bibitem[{{Fontaine} \& {Michaud}(1979)}]{Fontaine.Michaud79}
{Fontaine}, G. \& {Michaud}, G. 1979, \apj, 231, 826

\bibitem[{{Fontaine} {et~al.}(1984){Fontaine}, {Villeneuve}, {Wesemael}, \&
  {Wegner}}]{Fontaine.Villeneuve.ea84}
{Fontaine}, G., {Villeneuve}, B., {Wesemael}, F., \& {Wegner}, G. 1984, \apjl,
  277, L61

\bibitem[{{Fressin} {et~al.}(2013){Fressin}, {Torres}, {Charbonneau}, {Bryson},
  {Christiansen}, {Dressing}, {Jenkins}, {Walkowicz}, \&
  {Batalha}}]{fressin13-1}
{Fressin}, F., {Torres}, G., {Charbonneau}, D., {et~al.} 2013, \apj, 766, 81

\bibitem[{{Frewen} \& {Hansen}(2014)}]{Frewen.Hansen14}
{Frewen}, S.~F.~N. \& {Hansen}, B.~M.~S. 2014, \mnras, 439, 2442

\bibitem[{{Friedrich} {et~al.}(2004){Friedrich}, {Jordan}, \&
  {Koester}}]{Friedrich.Jordan.ea04}
{Friedrich}, S., {Jordan}, S., \& {Koester}, D. 2004, \aap, 424, 665

\bibitem[{{Gaidos}(2013)}]{Gaidos13}
{Gaidos}, E. 2013, \apj, 770, 90

\bibitem[{{G{\"a}nsicke} {et~al.}(2012){G{\"a}nsicke}, {Koester}, {Farihi},
  {Girven}, {Parsons}, \& {Breedt}}]{Gansicke.Koester.ea12}
{G{\"a}nsicke}, B.~T., {Koester}, D., {Farihi}, J., {et~al.} 2012, \mnras, 424,
  333

\bibitem[{{G{\"a}nsicke} {et~al.}(2008){G{\"a}nsicke}, {Koester}, {Marsh},
  {Rebassa-Mansergas}, \& {Southworth}}]{Gansicke.Koester.ea08}
{G{\"a}nsicke}, B.~T., {Koester}, D., {Marsh}, T.~R., {Rebassa-Mansergas}, A.,
  \& {Southworth}, J. 2008, \mnras, 391, L103

\bibitem[{{G{\"a}nsicke} {et~al.}(2007){G{\"a}nsicke}, {Marsh}, \&
  {Southworth}}]{Gansicke.Marsh.ea07}
{G{\"a}nsicke}, B.~T., {Marsh}, T.~R., \& {Southworth}, J. 2007, \mnras, 380,
  L35

\bibitem[{{G{\"a}nsicke} {et~al.}(2006){G{\"a}nsicke}, {Marsh}, {Southworth},
  \& {Rebassa-Mansergas}}]{Gansicke.Marsh.ea06}
{G{\"a}nsicke}, B.~T., {Marsh}, T.~R., {Southworth}, J., \&
  {Rebassa-Mansergas}, A. 2006, Science, 314, 1908

\bibitem[{{Giammichele} {et~al.}(2012){Giammichele}, {Bergeron}, \&
  {Dufour}}]{giammicheleetal12-1}
{Giammichele}, N., {Bergeron}, P., \& {Dufour}, P. 2012, \apjs, 199, 29

\bibitem[{{Gianninas} {et~al.}(2011){Gianninas}, {Bergeron}, \&
  {Ruiz}}]{Gianninas.Bergeron.ea11}
{Gianninas}, A., {Bergeron}, P., \& {Ruiz}, M.~T. 2011, \apj, 743, 138

\bibitem[{{Girven} {et~al.}(2012){Girven}, {Brinkworth}, {Farihi},
  {G{\"a}nsicke}, {Hoard}, {Marsh}, \& {Koester}}]{girvenetal12-1}
{Girven}, J., {Brinkworth}, C.~S., {Farihi}, J., {et~al.} 2012, \apj, 749, 154

\bibitem[{{Girven} {et~al.}(2011){Girven}, {G{\"a}nsicke}, {Steeghs}, \&
  {Koester}}]{girvenetal11-1}
{Girven}, J., {G{\"a}nsicke}, B.~T., {Steeghs}, D., \& {Koester}, D. 2011,
  \mnras, 417, 1210

\bibitem[{{Gonzalez} {et~al.}(1995){Gonzalez}, {LeBlanc}, {Artru}, \&
  {Michaud}}]{Gonzalez.LeBlanc.ea95}
{Gonzalez}, J.-F., {LeBlanc}, F., {Artru}, M.-C., \& {Michaud}, G. 1995, \aap,
  297, 223

\bibitem[{{Graham} {et~al.}(1990){Graham}, {Matthews}, {Neugebauer}, \&
  {Soifer}}]{Graham.Matthews.ea90}
{Graham}, J.~R., {Matthews}, K., {Neugebauer}, G., \& {Soifer}, B.~T. 1990,
  \apj, 357, 216

\bibitem[{{Green} {et~al.}(2012){Green}, {Froning}, {Osterman}, {Ebbets},
  {Heap}, {Leitherer}, {Linsky}, {Savage}, {Sembach}, {Shull}, {Siegmund},
  {Snow}, {Spencer}, {Stern}, {Stocke}, {Welsh}, {B{\'e}land}, {Burgh},
  {Danforth}, {France}, {Keeney}, {McPhate}, {Penton}, {Andrews},
  {Brownsberger}, {Morse}, \& {Wilkinson}}]{greenetal12-1}
{Green}, J.~C., {Froning}, C.~S., {Osterman}, S., {et~al.} 2012, \apj, 744, 60

\bibitem[{{Hartmann} {et~al.}(2011){Hartmann}, {Nagel}, {Rauch}, \&
  {Werner}}]{Hartmann.Nagel.ea11}
{Hartmann}, S., {Nagel}, T., {Rauch}, T., \& {Werner}, K. 2011, \aap, 530, A7

\bibitem[{{Johnson} {et~al.}(2010){Johnson}, {Aller}, {Howard}, \&
  {Crepp}}]{johnsonetal10-1}
{Johnson}, J.~A., {Aller}, K.~M., {Howard}, A.~W., \& {Crepp}, J.~R. 2010,
  \pasp, 122, 905

\bibitem[{{Johnson} {et~al.}(2013){Johnson}, {Morton}, \&
  {Wright}}]{johnsonetal13-1}
{Johnson}, J.~A., {Morton}, T.~D., \& {Wright}, J.~T. 2013, \apj, 763, 53

\bibitem[{{Jura}(2003)}]{Jura03}
{Jura}, M. 2003, \apjl, 584, L91

\bibitem[{{Jura}(2008)}]{Jura08}
{Jura}, M. 2008, \aj, 135, 1785

\bibitem[{{Jura} {et~al.}(2007){Jura}, {Farihi}, \&
  {Zuckerman}}]{Jura.Farihi.ea07}
{Jura}, M., {Farihi}, J., \& {Zuckerman}, B. 2007, \apj, 663, 1285

\bibitem[{{Jura} {et~al.}(2012){Jura}, {Xu}, {Klein}, {Koester}, \&
  {Zuckerman}}]{juraetal12-1}
{Jura}, M., {Xu}, S., {Klein}, B., {Koester}, D., \& {Zuckerman}, B. 2012,
  \apj, 750, 69

\bibitem[{{Kalirai} {et~al.}(2008){Kalirai}, {Hansen}, {Kelson}, {Reitzel},
  {Rich}, \& {Richer}}]{kaliraietal08-1}
{Kalirai}, J.~S., {Hansen}, B.~M.~S., {Kelson}, D.~D., {et~al.} 2008, \apj,
  676, 594

\bibitem[{{Kilic} {et~al.}(2012){Kilic}, {Patterson}, {Barber}, {Leggett}, \&
  {Dufour}}]{kilicetal12-1}
{Kilic}, M., {Patterson}, A.~J., {Barber}, S., {Leggett}, S.~K., \& {Dufour},
  P. 2012, \mnras, 419, L59

\bibitem[{{Klein} {et~al.}(2011){Klein}, {Jura}, {Koester}, \&
  {Zuckerman}}]{Klein.Jura.ea11}
{Klein}, B., {Jura}, M., {Koester}, D., \& {Zuckerman}, B. 2011, \apj, 741, 64

\bibitem[{{Klein} {et~al.}(2010){Klein}, {Jura}, {Koester}, {Zuckerman}, \&
  {Melis}}]{Klein.Jura.ea10}
{Klein}, B., {Jura}, M., {Koester}, D., {Zuckerman}, B., \& {Melis}, C. 2010,
  \apj, 709, 950

\bibitem[{{Koester}(2009)}]{Koester09}
{Koester}, D. 2009, \aap, 498, 517

\bibitem[{{Koester}(2010)}]{Koester10}
{Koester}, D. 2010, \memsai, 81, 921

\bibitem[{{Koester} {et~al.}(2005){Koester}, {Rollenhagen}, {Napiwotzki},
  {Voss}, {Christlieb}, {Homeier}, \& {Reimers}}]{Koester.Rollenhagen.ea05*b}
{Koester}, D., {Rollenhagen}, K., {Napiwotzki}, R., {et~al.} 2005, \aap, 432,
  1025

\bibitem[{{Koester} {et~al.}(2009){Koester}, {Voss}, {Napiwotzki},
  {Christlieb}, {Homeier}, {Lisker}, {Reimers}, \& {Heber}}]{Koester.Voss.ea09}
{Koester}, D., {Voss}, B., {Napiwotzki}, R., {et~al.} 2009, \aap, 505, 441

\bibitem[{{Koester} {et~al.}(1982){Koester}, {Weidemann}, \&
  {Zeidler-KT}}]{Koester.Weidemann.ea82}
{Koester}, D., {Weidemann}, V., \& {Zeidler-KT}, E.~M. 1982, \aap, 116, 147

\bibitem[{{Koester} \& {Wilken}(2006)}]{Koester.Wilken06}
{Koester}, D. \& {Wilken}, D. 2006, \aap, 453, 1051

\bibitem[{{Liebert} {et~al.}(2005){Liebert}, {Bergeron}, \&
  {Holberg}}]{Liebert.Bergeron.ea05}
{Liebert}, J., {Bergeron}, P., \& {Holberg}, J.~B. 2005, \apjs, 156, 47

\bibitem[{{Lloyd}(2011)}]{lloyd11-1}
{Lloyd}, J.~P. 2011, \apjl, 739, L49

\bibitem[{{Lloyd}(2013)}]{lloyd13-1}
{Lloyd}, J.~P. 2013, \apjl, 774, L2

\bibitem[{{Marsh} {et~al.}(1995){Marsh}, {Dhillon}, \&
  {Duck}}]{Marsh.Dhillon.ea95}
{Marsh}, T.~R., {Dhillon}, V.~S., \& {Duck}, S.~R. 1995, \mnras, 275, 828

\bibitem[{{Maxted} \& {Marsh}(1998)}]{Maxted.Marsh98}
{Maxted}, P. F.~L. \& {Marsh}, T.~R. 1998, \mnras, 296, L34

\bibitem[{{Maxted} {et~al.}(2000){Maxted}, {Marsh}, \&
  {North}}]{Maxted.Marsh.ea00}
{Maxted}, P.~F.~L., {Marsh}, T.~R., \& {North}, R.~C. 2000, \mnras, 317, L41

\bibitem[{{McDonough}(2000)}]{mcdonough00-1}
{McDonough}, W. 2000, in Earthquake Thermodynamics and Phase Transformation in
  the Earth's Interior, ed. R.~Teisseyre \& E.~Majewski (Elsevier Science
  Academic Press), 5--24

\bibitem[{{Melis} {et~al.}(2012){Melis}, {Dufour}, {Farihi}, {Bochanski},
  {Burgasser}, {Parsons}, {G{\"a}nsicke}, {Koester}, \&
  {Swift}}]{melisetal12-1}
{Melis}, C., {Dufour}, P., {Farihi}, J., {et~al.} 2012, \apjl, 751, L4

\bibitem[{{Metzger} {et~al.}(2012){Metzger}, {Rafikov}, \&
  {Bochkarev}}]{metzgeretal12-1}
{Metzger}, B.~D., {Rafikov}, R.~R., \& {Bochkarev}, K.~V. 2012, \mnras, 423,
  505

\bibitem[{{Michaud} {et~al.}(1979){Michaud}, {Martel}, {Montmerle}, {Cox},
  {Magee}, \& {Hodson}}]{Michaud.Martel.ea79}
{Michaud}, G., {Martel}, A., {Montmerle}, T., {et~al.} 1979, \apj, 234, 206

\bibitem[{{Mustill} \& {Villaver}(2012)}]{mustill+villaver12-1}
{Mustill}, A.~J. \& {Villaver}, E. 2012, \apj, 761, 121

\bibitem[{{Napiwotzki} {et~al.}(2001){Napiwotzki}, {Christlieb}, {Drechsel},
  {Hagen}, {Heber}, {Homeier}, {Karl}, {Koester}, {Leibundgut}, {Marsh},
  {Moehler}, {Nelemans}, {Pauli}, {Reimers}, {Renzini}, \&
  {Yungelson}}]{Napiwotzki.Christlieb.ea01}
{Napiwotzki}, R., {Christlieb}, N., {Drechsel}, H., {et~al.} 2001,
  Astronomische Nachrichten, 322, 411

\bibitem[{{Nelemans} {et~al.}(2005){Nelemans}, {Napiwotzki}, {Karl}, {Marsh},
  {Voss}, {Roelofs}, {Izzard}, {Montgomery}, {Reerink}, {Christlieb}, \&
  {Reimers}}]{Nelemans.Napiwotzki.ea05}
{Nelemans}, G., {Napiwotzki}, R., {Karl}, C., {et~al.} 2005, \aap, 440, 1087

\bibitem[{{Nelemans} \& {Tauris}(1998)}]{Nelemans.Tauris98}
{Nelemans}, G. \& {Tauris}, T.~M. 1998, \aap, 335, L85

\bibitem[{{Nordhaus} {et~al.}(2010){Nordhaus}, {Spiegel}, {Ibgui}, {Goodman},
  \& {Burrows}}]{nordhausetal10-1}
{Nordhaus}, J., {Spiegel}, D.~S., {Ibgui}, L., {Goodman}, J., \& {Burrows}, A.
  2010, \mnras, 408, 631

\bibitem[{{Paquette} {et~al.}(1986){Paquette}, {Pelletier}, {Fontaine}, \&
  {Michaud}}]{Paquette.Pelletier.ea86}
{Paquette}, C., {Pelletier}, C., {Fontaine}, G., \& {Michaud}, G. 1986, \apjs,
  61, 177

\bibitem[{{Pelletier} {et~al.}(1986){Pelletier}, {Fontaine}, {Wesemael},
  {Michaud}, \& {Wegner}}]{Pelletier.Fontaine.ea86}
{Pelletier}, C., {Fontaine}, G., {Wesemael}, F., {Michaud}, G., \& {Wegner}, G.
  1986, \apj, 307, 242

\bibitem[{{Rafikov}(2011{\natexlab{a}})}]{rafikov11-1}
{Rafikov}, R.~R. 2011{\natexlab{a}}, \apjl, 732, L3

\bibitem[{{Rafikov}(2011{\natexlab{b}})}]{rafikov11-2}
{Rafikov}, R.~R. 2011{\natexlab{b}}, \mnras, 416, L55

\bibitem[{{Rafikov} \& {Garmilla}(2012)}]{rafikov+garmilla12-1}
{Rafikov}, R.~R. \& {Garmilla}, J.~A. 2012, \apj, 760, 123

\bibitem[{{Rebassa-Mansergas} {et~al.}(2011){Rebassa-Mansergas}, {Nebot
  G{\'o}mez-Mor{\'a}n}, {Schreiber}, {Girven}, \&
  {G{\"a}nsicke}}]{rebassa-mansergasetal11-1}
{Rebassa-Mansergas}, A., {Nebot G{\'o}mez-Mor{\'a}n}, A., {Schreiber}, M.~R.,
  {Girven}, J., \& {G{\"a}nsicke}, B.~T. 2011, \mnras, 413, 1121

\bibitem[{{Redfield} \& {Linsky}(2004)}]{Redfield.Linsky04}
{Redfield}, S. \& {Linsky}, J.~L. 2004, \apj, 602, 776

\bibitem[{{Sackmann} {et~al.}(1993){Sackmann}, {Boothroyd}, \&
  {Kraemer}}]{Sackmann.Boothroyd.ea93}
{Sackmann}, I.~J., {Boothroyd}, A.~I., \& {Kraemer}, K.~E. 1993, \apj, 418, 457

\bibitem[{{Santos} \& {Kepler}(2012)}]{Santos.Kepler12}
{Santos}, M.~G. \& {Kepler}, S.~O. 2012, \mnras, 423, 68

\bibitem[{{Savage} \& {Sembach}(1996)}]{Savage.Sembach96}
{Savage}, B.~D. \& {Sembach}, K.~R. 1996, \araa, 34, 279

\bibitem[{{Schatzman}(1947)}]{Schatzman47}
{Schatzman}, E. 1947, Annales d'Astrophysique, 10, 19

\bibitem[{{Schreiber} \& {G{\"a}nsicke}(2003)}]{schreiber+gaensicke03-1}
{Schreiber}, M.~R. \& {G{\"a}nsicke}, B.~T. 2003, \aa, 406, 305

\bibitem[{{Shipman} {et~al.}(1995){Shipman}, {Provencal}, {Roby}, {Barstow},
  {Bond}, {Bruhweiler}, {Finley}, {Fontaine}, {Holberg}, {Nousek}, {Sion},
  {Tweedy}, {Wesemael}, \& {Vauclair}}]{Shipman.Provencal.ea95}
{Shipman}, H.~L., {Provencal}, J., {Roby}, S.~W., {et~al.} 1995, \aj, 109, 1220

\bibitem[{{Steele} {et~al.}(2011){Steele}, {Burleigh}, {Dobbie}, {Jameson},
  {Barstow}, \& {Satterthwaite}}]{steeleetal11-1}
{Steele}, P.~R., {Burleigh}, M.~R., {Dobbie}, P.~D., {et~al.} 2011, \mnras,
  416, 2768

\bibitem[{{Su} {et~al.}(2007){Su}, {Chu}, {Rieke}, {Huggins}, {Gruendl},
  {Napiwotzki}, {Rauch}, {Latter}, \& {Volk}}]{suetal07-1}
{Su}, K.~Y.~L., {Chu}, Y., {Rieke}, G.~H., {et~al.} 2007, \apjl, 657, L41

\bibitem[{{Tremblay} \& {Bergeron}(2009)}]{Tremblay.Bergeron09}
{Tremblay}, P.-E. \& {Bergeron}, P. 2009, \apj, 696, 1755

\bibitem[{{Vauclair} {et~al.}(1979){Vauclair}, {Vauclair}, \&
  {Greenstein}}]{Vauclair.Vauclair.ea79}
{Vauclair}, G., {Vauclair}, S., \& {Greenstein}, J.~L. 1979, \aap, 80, 79

\bibitem[{{Vennes} {et~al.}(2010){Vennes}, {Kawka}, \&
  {N{\'e}meth}}]{Vennes.Kawka.ea10}
{Vennes}, S., {Kawka}, A., \& {N{\'e}meth}, P. 2010, \mnras, 404, L40

\bibitem[{{Vennes} {et~al.}(2011{\natexlab{a}}){Vennes}, {Kawka}, \&
  {N{\'e}meth}}]{Vennes.Kawka.ea11}
{Vennes}, S., {Kawka}, A., \& {N{\'e}meth}, P. 2011{\natexlab{a}}, \mnras, 413,
  2545

\bibitem[{{Vennes} {et~al.}(1988){Vennes}, {Pelletier}, {Fontaine}, \&
  {Wesemael}}]{Vennes.Pelletier.ea88}
{Vennes}, S., {Pelletier}, C., {Fontaine}, G., \& {Wesemael}, F. 1988, \apj,
  331, 876

\bibitem[{{Vennes} {et~al.}(2011{\natexlab{b}}){Vennes}, {Thorstensen},
  {Kawka}, {N{\'e}meth}, {Skinner}, {Pigulski}, {Ste{\c}e{\'s}licki},
  {Ko{\l}aczkowski}, \& {{\'S}r{\'o}dka}}]{vennesetal11-1}
{Vennes}, S., {Thorstensen}, J.~R., {Kawka}, A., {et~al.} 2011{\natexlab{b}},
  \apjl, 737, L16+

\bibitem[{{Veras} {et~al.}(2013){Veras}, {Mustill}, {Bonsor}, \&
  {Wyatt}}]{Veras.Mustill.ea13}
{Veras}, D., {Mustill}, A.~J., {Bonsor}, A., \& {Wyatt}, M.~C. 2013, \mnras,
  431, 1686

\bibitem[{{Veras} {et~al.}(2011){Veras}, {Wyatt}, {Mustill}, {Bonsor}, \&
  {Eldridge}}]{verasetal11-1}
{Veras}, D., {Wyatt}, M.~C., {Mustill}, A.~J., {Bonsor}, A., \& {Eldridge},
  J.~J. 2011, \mnras, 417, 2104

\bibitem[{{Villaver} \& {Livio}(2007)}]{villaver+livio07-1}
{Villaver}, E. \& {Livio}, M. 2007, \apj, 661, 1192

\bibitem[{{Villaver} \& {Livio}(2009)}]{villaver+livo09-1}
{Villaver}, E. \& {Livio}, M. 2009, \apjl, 705, L81

\bibitem[{{von Hippel} {et~al.}(2007){von Hippel}, {Kuchner}, {Kilic},
  {Mullally}, \& {Reach}}]{von-Hippel.Kuchner.ea07}
{von Hippel}, T., {Kuchner}, M.~J., {Kilic}, M., {Mullally}, F., \& {Reach},
  W.~T. 2007, \apj, 662, 544

\bibitem[{{Voyatzis} {et~al.}(2013){Voyatzis}, {Hadjidemetriou}, {Veras}, \&
  {Varvoglis}}]{voyatzisetal13-1}
{Voyatzis}, G., {Hadjidemetriou}, J.~D., {Veras}, D., \& {Varvoglis}, H. 2013,
  \mnras, 430, 3383

\bibitem[{{Weidemann}(1984)}]{Weidemann84}
{Weidemann}, V. 1984, in IAU Symp. 105: Observational Tests of the Stellar
  Evolution Theory, Vol. 105, 229

\bibitem[{{Weidemann}(2000)}]{Weidemann00}
{Weidemann}, V. 2000, \aap, 363, 647

\bibitem[{{Williams} {et~al.}(2009){Williams}, {Bolte}, \&
  {Koester}}]{Williams.Bolte.ea09}
{Williams}, K.~A., {Bolte}, M., \& {Koester}, D. 2009, \apj, 693, 355

\bibitem[{{Wolff} {et~al.}(2002){Wolff}, {Koester}, \&
  {Liebert}}]{Wolff.Koester.ea02}
{Wolff}, B., {Koester}, D., \& {Liebert}, J. 2002, \aap, 385, 995

\bibitem[{{Xu} \& {Jura}(2012)}]{xu+jura12-1}
{Xu}, S. \& {Jura}, M. 2012, \apj, 745, 88

\bibitem[{{Xu} {et~al.}(2013{\natexlab{a}}){Xu}, {Jura}, {Klein}, {Koester}, \&
  {Zuckerman}}]{Xu.Jura.ea13}
{Xu}, S., {Jura}, M., {Klein}, B., {Koester}, D., \& {Zuckerman}, B.
  2013{\natexlab{a}}, \apj, 766, 132

\bibitem[{{Xu} {et~al.}(2013{\natexlab{b}}){Xu}, {Jura}, {Koester}, {Klein}, \&
  {Zuckerman}}]{xuetal13-2}
{Xu}, S., {Jura}, M., {Koester}, D., {Klein}, B., \& {Zuckerman}, B.
  2013{\natexlab{b}}, \apjl, 766, L18

\bibitem[{{Xu} {et~al.}(2014){Xu}, {Jura}, {Koester}, {Klein}, \&
  {Zuckerman}}]{Xu.Jura.ea14}
{Xu}, S., {Jura}, M., {Koester}, D., {Klein}, B., \& {Zuckerman}, B. 2014,
  \apj, 783, 79

\bibitem[{{Zuckerman} \& {Becklin}(1987)}]{Zuckerman.Becklin87}
{Zuckerman}, B. \& {Becklin}, E.~E. 1987, \apjl, 319, L99

\bibitem[{{Zuckerman} {et~al.}(2011){Zuckerman}, {Koester}, {Dufour}, {Melis},
  {Klein}, \& {Jura}}]{Zuckerman.Koester.ea11}
{Zuckerman}, B., {Koester}, D., {Dufour}, P., {et~al.} 2011, \apj, 739, 101

\bibitem[{{Zuckerman} {et~al.}(2007){Zuckerman}, {Koester}, {Melis}, {Hansen},
  \& {Jura}}]{Zuckerman.Koester.ea07}
{Zuckerman}, B., {Koester}, D., {Melis}, C., {Hansen}, B.~M., \& {Jura}, M.
  2007, \apj, 671, 872

\bibitem[{{Zuckerman} {et~al.}(2003){Zuckerman}, {Koester}, {Reid}, \& {H{\"
  u}nsch}}]{Zuckerman.Koester.ea03}
{Zuckerman}, B., {Koester}, D., {Reid}, I.~N., \& {H{\" u}nsch}, M. 2003, \apj,
  596, 477

\bibitem[{{Zuckerman} {et~al.}(2010){Zuckerman}, {Melis}, {Klein}, {Koester},
  \& {Jura}}]{zuckermanetal10-1}
{Zuckerman}, B., {Melis}, C., {Klein}, B., {Koester}, D., \& {Jura}, M. 2010,
  \apj, 722, 725

\bibitem[{{Zuckerman} {et~al.}(2013){Zuckerman}, {Xu}, {Klein}, \&
  {Jura}}]{zuckermanetal13-1}
{Zuckerman}, B., {Xu}, S., {Klein}, B., \& {Jura}, M. 2013, \apj, 770, 140

\end{thebibliography}
\end{document}